\documentclass[twocolumn,aps,epsfig,nofootinbib,preprintnumbers]{revtex4-1}
\usepackage{graphicx}
\usepackage{epstopdf}
\usepackage{latexsym}
\usepackage{amssymb}
\usepackage{amsmath}
\usepackage{color}
\usepackage{float}
\usepackage{mathrsfs}
\usepackage{multirow}
\usepackage{tabularx,booktabs}
\usepackage{esvect}
\usepackage{ulem} 
\newcolumntype{Y}{>{\centering\arraybackslash}X}

\usepackage[center]{subfigure}
\usepackage{makecell}
\usepackage{ulem}
\usepackage[colorlinks=true]{hyperref} 
\setcellgapes{4pt}
\begin{document}

  \renewcommand\arraystretch{2}
 \newcommand{\bq}{\begin{equation}}
 \newcommand{\eq}{\end{equation}}
 \newcommand{\bqn}{\begin{eqnarray}}
 \newcommand{\eqn}{\end{eqnarray}}
 \newcommand{\nb}{\nonumber}
 \newcommand{\lb}{\label}
 \newcommand{\cb}{\color{blue}}
    \newcommand{\cc}{\color{cyan}}
        \newcommand{\cm}{\color{magenta}}
\newcommand{\rc}{\rho^{\scriptscriptstyle{\mathrm{I}}}_c}
\newcommand{\rd}{\rho^{\scriptscriptstyle{\mathrm{II}}}_c} 
\newcommand{\PRL}{Phys. Rev. Lett.}
\newcommand{\PL}{Phys. Lett.}
\newcommand{\PR}{Phys. Rev.}
\newcommand{\CQG}{Class. Quantum Grav.}
\newcommand{\delete}[1] {\red{\sout{#1}}}

\newcommand{\ny}[1]{\textcolor{blue}{\bf{ny: #1}}}
\newcommand{\ky}[1]{ {\bf{ky: #1}}}
\newcommand{\kent}[1]{ {#1}}


\title{Gravitational waves from the quasi-circular inspiral of compact binaries in Einstein-aether theory} 
 
\author{Chao Zhang$^{1}$}
 %
 \author{Xiang Zhao$^{1}$}
%
\author{Anzhong Wang$^{1}$ }
\email{Anzhong$\_$Wang@baylor.edu; Corresponding Author}
\author{Bin Wang$^{2, 3}$}
%
\author{Kent Yagi$^{4}$}
%
\author{Nicolas Yunes$^5$}
%
\author{Wen Zhao$^{6,7}$}
%
\author{Tao Zhu$^8$}
%

\affiliation{$^{1}$ GCAP-CASPER, Physics Department, Baylor University, Waco, TX, 76798-7316, USA}
\affiliation{$^{2}$ Center for Gravitation and Cosmology, Yangzhou University, Yangzhou 225009, China}
\affiliation{$^{3}$ School of Aeronautics and Astronautics, Shanghai Jiao Tong University, Shanghai 200240, China}
\affiliation{$^{4}$ Department of Physics, University of Virginia, Charlottesville, VA, 22904, United States}
\affiliation{$^{5}$ Department of Physics, University of Illinois at Urbana-Champaign, Urbana, IL, 61801, United States}
 \affiliation{$^{6}$ CAS Key Laboratory for Researches in Galaxies and Cosmology, Department of Astronomy, \\
University of Science and Technology of China, Chinese Academy of Sciences, Hefei, Anhui 230026, China}
\affiliation{$^{7}$ School of Astronomy and Space Science, University of Science and Technology of China, Hefei 230026, China}
\affiliation{$^{8}$ {Institute for Theoretical Physics $\&$ Cosmology, Zhejiang University of Technology,} Hangzhou 310032, China}

\date{\today}

\begin{abstract}

We study gravitational waves emitted by a binary system of non-spinning bodies in a quasi-circular inspiral within the framework of Einstein-aether theory. 
In particular, we compute explicitly and analytically the expressions for the time-domain  and frequency-domain waveforms, gravitational wave polarizations, 
and response functions for both ground- and space-based detectors in the post-Newtonian approximation. We find that, when going beyond leading-order in 
the post-Newtonian approximation, the non-Einsteinian polarization modes contain terms that depend on both the first and the second harmonics of the orbital 
phase. We also calculate analytically  the corresponding parameterized post-Einsteinian parameters, generalizing the existing framework to allow for different 
propagation speeds among scalar, vector and tensor modes, without assumptions about the magnitude of its coupling parameters, and meanwhile allowing 
the binary system to have relative motions with respect to the aether field. Such results allow for the easy construction of Einstein-aether templates that could 
be used in Bayesian tests of General Relativity in the future.

 \end{abstract}


\maketitle

\section{Introduction}
 \renewcommand{\theequation}{1.\arabic{equation}} \setcounter{equation}{0}

The detection of the first gravitational wave (GW) from the coalescence of two massive black holes (BHs) by advanced LIGO marked the beginning of the GW era \cite{Ref1}. Following this observation, a few tens of GW candidates were identified by the LIGO/Virgo scientific collaboration \cite{GWs}\footnote{Recently, various GWs have been detected after LIGO/Virgo resumed operations on April 1, 2019, possibly including the coalescence of a neutron-star (NS)/BH binary. The details of these detections have not yet been released \cite{LIGO}.}. The LIGO and Virgo detectors are sensitive to GWs with frequencies between $20$ and $2000$Hz \cite{Ref3}, since at frequencies lower than $20$Hz they are limited by the Newtonian ground noise. As a consequence, LIGO and Virgo are only able to observe GWs produced in the late inspiral and merger of low-mass compact binaries, such as binary black holes (BBHs), BH-NSs, and binary neutron stars (BNSs). 
 
One of the many remarkable observational results obtained so far  is the discovery that the BBHs can be composed of objects with individual masses much larger than what was previously expected, both theoretically and observationally \cite{Ref4,Ref5,Ref6}, leading to the proposal and refinement of various formation scenarios \cite{Ref7,Ref8}. A consequence of this discovery is that the early inspiral phase may also be detectable by space-based observatories, such as LISA, TianQin, Taiji and DECIGO, for several years prior to their coalescence \cite{AS16, Moore15}. The analysis of the BBHs' population observed by LIGO and Virgo has shown that such space-based detectors may be able to see many such systems, with a variety of profound scientific consequences.
 
In particular, multiple observations with different detectors at different frequencies of signals from the same source can provide excellent opportunities to study the evolution of the binary in detail. Since different detectors observe at disjoint frequency bands, together they cover different evolutionary stages of the same binary system. Each stage of the evolution carries information about different physical aspects of the source. Technically, it also provides early warnings for an upcoming coalescence, so that ground-based detectors could know the sky localization of the source and its time to coalescence well in advance. 
  
Combining high- and low-frequency GW detections of the same source can also help identify the astrophysical channel responsible of BBH formations. Different scenarios in fact result in different masses, mass ratios, spins and eccentricity distributions of the  {detected sources \cite{Ref12,Ref13, Davide17, Davide18, Ben18, Michael17}.}  Because of the GW circularization, BBHs may have small eccentricity in the LIGO/Virgo band, regardless of their formation channels. However, space-based detectors will be able to observe GW signals from BBHs that did not have enough time to fully circularize, allowing for measurements of eccentricities in excess of $10^{-3}$~\cite{Ref13}. In addition, stellar-mass BBHs observed in the space-based detector bands provide a very promising class of  standard sirens (see, e.g., \cite{Holz18}). In the absence of a distinctive electromagnetic counterpart, it was estimated \cite{Ref14} that LISA might measure the Hubble constant within a few percent error, thus helping in the resolution of the discrepancy between the local measurement of this quantity \cite{Ref15}  and that obtained from the cosmic microwave background (CMB) \cite{Ref16}  {(Note that using ground-based detectors, e.g., aLIGO, the Hubble constant could also be measured with good precisions even if we do not identify electromagnetic counterparts  \cite{Nishi17, Chen16})}.
  
In addition, multi-band GW detections will enhance the potential to test gravitational theories in the strong, dynamical field regime of merging compact objects \cite{Ref17,Carson:2019rda,Carson:2019fxr,Carson:2019yxq,Gnocchi:2019jzp,Carson:2019kkh}. Massive systems will be observed by ground-based  detectors with high signal-to-noise ratios, after being tracked for years by space-based detectors in their inspiral phase. The two portions of signals can be combined to make precise tests for different theories of gravity. In particular, joint observations of BBHs with a total mass larger than about $60$ solar masses by LIGO/Virgo and space-based detectors can potentially improve current bounds on dipole emission from BBHs by more than six orders of magnitude \cite{Ref17}, which will impose severe constraints on various theories of gravity \cite{Ref18}. 
  
All the above work, however, depends crucially on the accurate description of GWs in order to track the signal during the early inspiral phase all the way to the merger phase. During the inspiral phase, GWs can be modeled using the post-Newtonian (PN) formalism \cite{Ref21}. Within general relativity (GR), waveforms at low PN orders (i.e.,~at or below the 2PN order) are sufficiently accurate for an unbiased recovery of the source parameters \cite{Ref22}. As the signal-to-noise ratio increases, however, our ability to test GR will be systematically limited by the accuracy of our waveform models. 
   
In recent work, some of the present authors generalized the PN formalism to certain modified theories of gravity and applied it to the quasi-circular inspiral of compact binaries. In particular, we calculated in detail the waveforms, gravitational wave polarizations, response functions and energy losses due to gravitational radiation in Brans-Dicke (BD) theory \cite{Ref23} and screened modified gravity (SMG) \cite{Tan18,Ref25,Ref25b} to the leading PN order, with which we then considered projected constraints from the third-generation detectors. Such studies have been further generalized to triple systems in Einstein-aether theory \cite{Kai19,Zhao19}. When applying such formulas to the first relativistic triple system discovered in 2014 \cite{Ransom14}, we studied the radiation power, and found that quadrupole emission has almost the same amplitude as that in GR, but the dipole emission can be as big as the quadrupole emission. This can provide a promising window to place severe constraints on Einstein-aether theory with multi-band gravitational wave observations \cite{Ref17,Carson:2019yxq}. 

In this paper, we study the gravitational waves emitted by a compact binary during its quasi-circular inspiral within Einstein-aether theory. This is, of course, not the first time gravitational waves that have been studied in this theory. The first studies were carried out by Foster in the mid 2000s \cite{Foster06,Foster07}, who computed the gravitational waves and the radiative losses of a generic binary through a multipolar decomposition. Using these results, Yagi, et al. \cite{Yagi:2013qpa,Yagi14} calculated the effects of such waves on the rate of change of the orbital period of binary pulsars, placing stringent constraints on a sector of the theory. Following this work, Hansen, et al.~\cite{HYY15} calculated the GW polarizations and response function in the time- and frequency-domain for a compact binary during its quasi-circular inspiral, but again in a restricted sector of the theory. More recently, more severe constraints were placed on Einstein-aether theory  \cite{Oost18,GHLP18}, using the recent binary NS observation by LIGO, which constrained the speed of gravity to better than one part in $10^{15}$~\cite{HYY15}.  

We here revisit some of these calculations without imposing any restrictions on the parameter space. First, we compute, once more, the gravitational waves emitted by a binary system and its associated radiative energy loss for a generic binary system in the PN approximation without assumptions about the magnitude of its coupling parameters. We then specialize this calculation to a compact binary in a quasi-circular inspiral and compute the time-domain response function both for ground- and space-based detectors. In doing so, we discover that previous expressions for the GW polarizations that compose the time-domain response function~\cite{Chat12} are not applicable to Einstein-aether theory due to the different speeds of propagation of its scalar and vector modes. This implies that the results of~\cite{HYY15} are corrected by terms that depend on these different speeds; in particular, this generates terms in the non-Einsteinian polarizations that depend explicitly on the speed of the center of mass of the binary with respect to the aether field. With these waveforms computed, we then calculate their Fourier transform in the stationary phase approximation (SPA) \cite{YP09,YBW09,Chat12}, and map the results to the parameterized post-Einsteinian (ppE) framework~\cite{YP09} that was extended to allow for different propagation speeds among different polarization modes~\cite{Chat12}.  Our results, therefore, allow for the straightforward construction of waveform templates with which to carry out tests of Einstein-aether theory using Bayesian theory and matched filtering in the future. 

The remainder of this paper presents the results summarized above. 
In particular, in Sec.~II we give a brief introduction to Einstein-aether theory  ($\ae$-theory for short),  and in Sec. III we calculate the GW polarizations and  energy loss rate, and correct some typos in the literature.
In Sec. IV, we  study the GW polarizations and response function for an inspiraling binary.
In Sec. V, we calculate the response function and its Fourier transform for both ground- and space-based detectors using the SPA \cite{YP09,YBW09,Chat12}.
In Sec. VI we map the results of the last section to the parametrized post-Einsteinian (ppE) framework~\cite{YP09,YBW09,Chat12},
while in Sec. VII, we summarize our main results and present discussions and concluding remarks. 
The paper also include four appendices, and in Appendix A, we present a brief review on the SPA, while in Appendices B, C and D we provide some additional mathematical formulas. 
We follow here the conventions of Misner, Thorne and Wheeler~\cite{MTW} and use units in which $c=1$. 

\section{Einstein-aether Theory}
 \renewcommand{\theequation}{2.\arabic{equation}} \setcounter{equation}{0}

In $\ae$-theory, the fundamental variables of the gravitational  sector are \cite{JM01}
\bq
\lb{var}
\left(g_{\mu\nu}, u^{\mu}, \lambda\right),
\eq
where $g_{\mu\nu}$ denotes the four-dimensional metric of spacetime
with signature $(-, +,+,+)$ \cite{Foster06,GEJ07},  $u^{\mu}$ the aether field, and $\lambda$ a Lagrangian multiplier that guarantees that the aether field is always timelike and unity,
\bq
\lb{con}
u^{\lambda} u_{\lambda} = -1.
\eq

{In this paper, we adopt the following conventions: all repeated Latin letters represent spatial indices that are to be summed over from $1$ to $3$, while repeated Greek letters represent 
spacetime indices to be summed over from $0$ to $3$, regardless of whether they are super-indices or sub-indices.
 
The general action of the theory  is given by  \cite{Jacobson},
\bq
\lb{act1}
S = S_{\ae} + S_{m},
\eq
where  $S_{m}$ denotes the matter action,  and $S_{\ae}$  is the gravitational action of $\ae$-theory,  given, respectively,  by
\bqn
\lb{act2}
S_{m} &=& \int{\sqrt{- g} \; d^4x \Big[{\cal{L}}_{m}\left(g_{\mu\nu}, u^{\alpha}; \psi_m\right)\Big]},  \\ \nb
 S_{\ae} &=& \frac{1}{16\pi G_{\ae} }\int{\sqrt{- g} \; d^4x \Big[R(g_{\mu\nu}) + {\cal{L}}_{\ae}\left(g_{\mu\nu}, u^{\alpha}, {\lambda}\right)\Big]}.
\eqn
Here $\psi_m$ collectively denotes the matter fields, $R$    and $g$ are, respectively, the  Ricci scalar and the determinant of $g_{\mu\nu}$,
 and
\bq
\lb{lag}
 {\cal{L}}_{\ae}  \equiv - M^{\alpha\beta}_{~~~~\mu\nu}\left(D_{\alpha}u^{\mu}\right) \left(D_{\beta}u^{\nu}\right) + \lambda \left(g_{\alpha\beta} u^{\alpha}u^{\beta} + 1\right),
\eq
 where $D_{\mu}$ denotes the covariant derivative with respect to $g_{\mu\nu}$. The tensor $M^{\alpha\beta}_{~~~~\mu\nu}$ is defined as
\bq
\lb{Mabcd}
M^{\alpha\beta}_{~~~~\mu\nu} \equiv c_1 g^{\alpha\beta} g_{\mu\nu} + c_2 \delta^{\alpha}_{\mu}\delta^{\beta}_{\nu} +  c_3 \delta^{\alpha}_{\nu}\delta^{\beta}_{\mu} - c_4 u^{\alpha}u^{\beta} g_{\mu\nu}.
\eq

Note that here we assume that matter fields couple not only to $g_{\mu\nu}$ but also to the aether field $u^{\mu}$,  in order 
to model effectively the radiation of a compact object  \cite{Foster07,Kai19}, such as a neutron star \cite{Ed75}. 
 {The current theoretical and observational 
constraints on the four dimensionless  coupling constants $c_i$'s were given explicitly in   \cite{Oost18}. It was found that
\bq
\lb{c1234a}
0 \le c_{14} \le 2.5 \times 10^{-5}, \quad \left|c_{13}\right| \le 10^{-15},
\eq
where $c_{ij}\equiv c_i +c_j$. The constraints on other parameters depend on the values of $c_{14}$. In particular, for $ 0 \lesssim c_{14}\leq 2\times 10^{-7}
$ and $ 2\times 10^{-6}\lesssim c_{14}\lesssim 2.5\times 10^{-5}$, they read  respectively as \cite{Oost18} (see also \cite{Sarbach:2019yso}),}
\begin{eqnarray}
\lb{c1234}
 && \mbox{(i)} \;\; 
  0 \lesssim c_{14}\leq 2\times 10^{-7}, \quad
   c_{14} \lesssim c_2 \lesssim  0.095, \nonumber\\
&&  \mbox{(ii)}\;\; 
  2\times 10^{-6}\lesssim c_{14}\lesssim 2.5\times 10^{-5}, \nonumber\\
 & & ~~~~~~ 0 \lesssim c_2-c_{14} \lesssim 2\times 10^{-7}.
\end{eqnarray}
In the intermediate regime of $c_{14}$ ($2\times 10^{-7}< c_{14}\lesssim 2\times 10^{-6}$), 
the results are obtained only numerically, and shown explicitly   in Fig.~1 of \cite{Oost18}.


 The coupling constant  $G_{\ae} $ is related to  the Newtonian gravitational constant $G_{N}$ via the relation \cite{CL04},
\bqn
\lb{Gae}
G_{N} = \frac{G_{\ae}}{1 - \frac{1}{2}c_{14}}.
\eqn

Strong field effects can be important  in the vicinity of a compact body, such as a neutron star or a black hole, and   need to be taken into account.
Following Eardley \cite{Ed75},  these effects  can be included by considering the test-particle action \cite{Foster07},
\bqn
\lb{act3}
S_A &=&  -\int{d\tau_A  \tilde{m}_A[ {\gamma_A}]} \nb\\
&=&  - \tilde{m}_A\int{d\tau_A \Bigg[1 + \sigma_A (1- {\gamma_A}) }\nb\\
&& + \frac{1}{2}(\sigma_A + \sigma_A^2 + \bar{\sigma}_A)(1- {\gamma_A})^2 + ...\Bigg],
\eqn
where  $ {\gamma_A}\equiv  -u^{\mu} v^A_{\mu}$,  and $v^A_{\mu}$ is the four-velocity of the body, defined as $v_A^i\equiv {dx_A^i}/{d\tau_A}$. 
The index $A$ labels the body, and $\tau_A$ is its  proper time. We also  note that  $\tilde{m}_A$ in (\ref{act3}) has the dimension of mass,   $\sigma_A$ and $\bar\sigma_A$ are defined as
\bqn
\lb{sen1}
\sigma_A  &\equiv& - \left.\frac{d\ln\tilde{m}_A[ {\gamma_A}]}{d\ln {\gamma_A}}\right|_{ {\gamma_A} = 1},\nb\\
\bar\sigma_A  &\equiv& { \left.\frac{d^2\ln\tilde{m}_A[ {\gamma_A}]}{d(\ln {\gamma_A})^2}\right|_{ {\gamma_A} = 1}},
\eqn
which can be determined by considering asymptotic properties of perturbations of static stellar configurations \cite{Yagi14}.

The variations of the total action with  respect to $g_{\mu\nu}$ and $u^{\mu}$ yield, respectively, the field equations \cite{Kai19}, 
 \bqn
 \lb{fie1}
 R^{\mu\nu} - \frac{1}{2} g^{\mu\nu}R - S^{\mu\nu} &=& 8\pi G_{\ae}  T^{\mu\nu},\\
 \lb{fie2}
 \AE_{\mu} &=& 8\pi G_{\ae}  T_{\mu},
\eqn
with the constraint of Eq.~(\ref{con}). Here we have \cite{Foster07},
 \bqn
 \lb{fie3}
  S_{\alpha\beta} &\equiv&
  D_{\mu}\Big[J^{\mu}{}_{(\alpha}u_{\beta)} + J_{(\alpha\beta)}u^{\mu}-u_{(\beta}J_{\alpha)}{}^{\mu}\Big]\nb\\
&& + c_1\Big[\left(D_{\alpha}u_{\mu}\right)\left(D_{\beta}u^{\mu}\right) - \left(D_{\mu}u_{\alpha}\right)\left(D^{\mu}u_{\beta}\right)\Big]\nb\\
&& + c_4 a_{\alpha}a_{\beta}    + \lambda  u_{\alpha}u_{\beta} - \frac{1}{2}  g_{\alpha\beta} J^{\delta}_{\;\;\sigma} D_{\delta}u^{\sigma},\nb\\
 \AE_{\mu} & \equiv &
 D_{\alpha} J^{\alpha}_{\;\;\;\mu} + c_4 a_{\alpha} D_{\mu}u^{\alpha} + \lambda u_{\mu},\nb\\
  T^{\mu\nu} &\equiv&  \frac{2}{\sqrt{-g}}\frac{\delta \left(\sqrt{-g} {\cal{L}}_{m}\right)}{\delta g_{\mu\nu}}\nb\\
           &=&\sum\limits_A \tilde{m}_A \tilde{\delta}_A [A_A^1 v_A^{\mu}v_A^{\nu}+2A_A^2 u^{(\mu}v_A^{\nu)}],\nb\\
T_{\mu} &\equiv& - \frac{1}{\sqrt{-g}}\frac{\delta \left(\sqrt{-g} {\cal{L}}_{m}\right)}{\delta u^{\mu}} 
= \sum\limits_A \tilde{m}_A \tilde{\delta}_A A_A^2 v_{\mu}^A, ~~~
 \eqn
 with  parentheses in index pairs denoting index symmetrization and
\begin{equation}
 \lb{Jab}
J^{\alpha}_{\;\;\;\mu} \equiv M^{\alpha\beta}_{~~~~\mu\nu}D_{\beta}u^{\nu}\,,\quad
a^{\mu} \equiv u^{\alpha}D_{\alpha}u^{\mu},
\end{equation}
and
\bqn
\lb{A12}
A_A^1&\equiv&1+\sigma_A+\frac{(\sigma_A + \sigma_A^2 + \bar{\sigma}_A)}{2}[(u_\mu v_A^\mu)^2-1],\nb\\
A_A^2&\equiv&-\sigma_A-(\sigma_A + \sigma_A^2 + \bar{\sigma}_A)(u_\mu v_A^\mu+1),\nb\\
\tilde{\delta}_A&\equiv&\frac{\delta^3(\vec{\bm{x}}-\vec{\bm{x}}_A)}{v_A^0 \sqrt{|g|}}.
\eqn
From Eq.~(\ref{fie2}) and the normalization condition, we also find that
\bq
\lb{lambda}
\lambda = u_{\beta}D_{\alpha}J^{\alpha\beta} + c_4 a^2 - 8\pi G_{\ae}  T_{\alpha}u^{\alpha},
\eq
where $a^{2}\equiv a_{\lambda}a^{\lambda}$.

\section{Gravitational Wave Polarizations and Energy Loss of Binary Systems in Einstein-aether theory}  
 \renewcommand{\theequation}{3.\arabic{equation}} \setcounter{equation}{0}

The linear perturbations of Einstein-aether theory over a Minkowski background were studied by several authors \cite{JM01,Foster06,Yagi14,Zhao19}. For the sake of convenience,
 in this section we first give a brief review of the relevant materials, following mostly \cite{Kai19}. For more  details on the PN approximations for many bodies in Einstein-aether theory, we refer the reader  to \cite{JM01,Foster06,Yagi14,Kai19}. Readers familiar with Einstein-aether theory may skip the first two subsections and go directly to the third section if they wish, in which we apply previous results to  binary system. 
 
Let us first note that 
\bq
\lb{M4}
g_{\mu\nu} = \eta_{\mu\nu}, \quad u^{\mu} = \delta^{\mu}_{t},
\eq
satisfies the Einstein-aether field equations in Eqs.~(\ref{fie1}) and (\ref{fie2}) in the coordinates $x^{\mu} = (t, x, y, z)$, where $\eta_{\mu\nu}= {\mbox{diag}}(-1, 1, 1, 1)$ is the Minkowski metric \cite{Invb}. Clearly, Eq.~(\ref{M4}) shows that the aether field $u^{\mu}$ is at rest in this Minkowski background \footnote{In cosmology, the aether field is often chosen to be 
comoving with the CMB \cite{Jacobson}. Thus, it is consistent here to choose the aether to be comoving with the Minkowski coordinate system $x^{\mu} = (t, x, y, z)$.}, so any motion with respect to this coordinate system also represents motion with respect to the aether field. In addition, as far as the aether field  is concerned, the time-like vector $u^{\mu}$ is invariant under the general spatial diffeomorphism ${x'}^j = {x'}^j\left(x^i\right), \; (i, j = 1, 2,  3)$. Later, without loss of generality,  we will use this gauge freedom to choose the plane of the binary system to coincide with the ($x, y$)-plane. 

Now, we consider the linear perturbations, 
 \bq
 \lb{per1}
h_{\mu\nu}=g_{\mu\nu}-\eta_{\mu\nu},~~~w^0=u^0-1,~~~w^i=u^i,
 \eq
where  $h_{\mu\nu}$, $w^0$ and $w^i$ are decomposed into the forms  \cite{Foster06},
 \bqn
 \lb{per2}
h_{0i}&=&\gamma_i+\gamma_{,i},~~~w_i=\nu_i+\nu_{,i},\nb\\
h_{ij}&=&\phi_{,ij}+\frac{1}{2}P_{ij}[f]  +2\phi_{(i,j)}+ \phi_{ij},
 \eqn
with $P_{ij} \equiv \delta_{ij}\Delta - \partial_i\partial_j$, where $ \Delta \equiv \delta^{ij}\partial_i \partial_j $.
In addition,  the vector and tensor fields  satisfy the conditions, 
 \bqn
 \lb{gau}
&& \partial^{i}\gamma_{i}= \partial^{i}\nu_{i}= \partial^{i}\phi_{i} = 0,  \nb\\
&&  \partial^{j}\phi_{ij}= 0, \quad \phi_{i}^{\;\; i}=0. 
\eqn
To the linear order in perturbation theory, it is convenient to define a non-symmetric tensor,
\bqn
\lb{tauab1}
\tau^{\mu\nu}\equiv T^{\mu\nu}-T^{\mu}\delta^{\nu}_{0},\;\;(\tau^{\mu\nu} \not= \tau^{\nu\mu}),
\eqn
which satisfies the conservation law   
\bqn
\lb{tauab2}
&&\partial^{\nu}\tau_{\mu\nu}=0.
\eqn
Defining the center-of-mass (COM) coordinate and its velocity as
\bqn
\lb{COMX}
&&X^i \equiv \frac{\sum\limits_A m_A x_A^i}{\sum\limits_A m_A}, \\
\lb{COMV}
&&V^i \equiv \frac{d X^i}{dt},
\eqn
we find that conservation of momentum requires
\bqn
\lb{COMV2}
\frac{dV^i}{dt}&=&0, \quad \Rightarrow \quad V^i = {\mbox{Constant}}.
\eqn

\subsection{Linearized Einstein-aether Field Equations}

Substituting the above expressions into the linearized Einstein-aether field equations, we find that the tensor, vector and scalar parts can be written as follows \cite{Kai19}.
For the tensor part,  we have
\bqn
\lb{spin2}
\frac{1}{c_T^2} \ddot{\phi}_{ij}-\Delta \phi_{ij}&=&16 \pi G_{\ae} \tau_{ij}^{TT},
\eqn
with
\bqn
\lb{cT}
c_T^2 &\equiv& \frac{1}{1- {c_{+}} },
\eqn
where {$c_+=c_{13}  \equiv c_1+c_3$}, and  ``TT"  stands for the transverse-traceless operator acting on the tensor.  

For the vector part,  we have \footnote{Notice that the last term of Eq.~\eqref{spin11} corrects a sign error in Eq.~(44) of \cite{Foster06}.}
\bqn
\lb{spin11}
 \frac{1}{c_V^2} (\ddot{\nu}^i+\ddot{\gamma}^i)- {\Delta (\nu^i+\gamma^i)} 
&=& \frac{16 \pi G_{\ae}}{2 c_1-c_{13} c_-} \nb\\ 
&& \times \left[c_{13}  \tau_{i0}-(1-c_{13} ) T^i\right]^T, \nb \\
\\
\lb{spin12}
\Delta\left(c_{13}  \nu_i+\gamma_i\right) &=& -16 \pi G_{\ae} \tau_{i0}^T,
\eqn
where
\bqn
\lb{cV}
c_V^2 \equiv \frac{2 c_1-c_{13}  c_-}{2 (1-c_{13} ) c_{14}},
\eqn
with $c_- \equiv c_1-c_3$,  and the T above stands for the transverse operator acting on the vector. 

For the scalar part,  we have
\bqn
\lb{spin01}
\frac{1}{c_S^2} \ddot{F}-\Delta F &=& \frac{16 \pi G_{\ae} c_{14}}{2-c_{14}} \left(\tau_{kk}+\frac{2}{c_{14}} \tau_{00}  \right. \nb\\
&& \left.  -\frac{2+3 c_2+c_{13} }{c_{123}} \tau_{kk}^L \right),\\
\lb{spin02}
 \Delta\left(F-c_{14} h_{00}\right) &=& -16 \pi G_{\ae} \tau_{00},\\
\lb{spin03}
 \left[(1+c_2) \dot{F} +c_{123} \Delta \dot{\phi}\right]_{,i}&=& -16 \pi G_{\ae} \tau_{i0}^L,
\eqn
where $F\equiv \Delta f$, and 
\bqn
\lb{cS}
c_S^2 \equiv \frac{(2-c_{14}) c_{123}}{(2+3 c_2+c_{13} ) (1-c_{13} ) c_{14}},
\eqn
with $c_{ijk} \equiv c_i+c_j+c_k$,  and the L above stands for the longitudinal operator acting on the vector.  
In addition, the constraint in Eq.~(\ref{con}) gives 
\bqn
\lb{u0}
 h_{00} = 2 w^0.
\eqn
From these equations, we can easily infer that the tensor, vector and scalar modes propagate with speeds $c_{T}$, $c_{V}$ and $c_{S}$, respectively.

\subsection{Gravitational Wave Polarizations and Energy Loss}

To consider the polarizations of gravitational waves in Einstein-aether theory, let us consider the time-like geodesic deviation equation. 
 In the spacetime described by the metric, $g_{\mu\nu} = \eta_{\mu\nu} + h_{\mu\nu}$,
 the spatial deviation vector, $\zeta_i$, satisfies
 \bqn
 \lb{3.22}
\ddot{\zeta}_i=-R_{0i0j}\zeta^j\equiv\frac{1}{2}\ddot{{\cal{P}}}_{ij}\zeta^j,
 \eqn
where $\zeta_{\mu}$ describes the four-dimensional deviation vector  between two nearby trajectories of test particles, and
  \bqn
 \lb{3.23}
R_{0i0j}&\simeq& \frac{1}{2}(h_{0j,0i}+h_{0i,0j}-h_{ij,00}-h_{00,ij})\nb\\
&=&-\frac{1}{2}\ddot{\phi}_{ij}+\dot{\Psi}^{\mathrm{II}}_{(i,j)}+\Phi^{\mathrm{IV}}_{,ij}-\frac{1}{2}\delta_{ij}\ddot{\Phi}^{\mathrm{II}},
 \eqn
where ${\Psi}^{\mathrm{II}}_{i}, \; \Phi^{\mathrm{IV}}$ and ${\Phi}^{\mathrm{II}}$ are the gauge-invariant quantities defined in \cite{Kai19}. In particular, we have
$\Phi^{\mathrm{II}} \equiv F/2$.

 In the wave zone, $|\vec{x}|  \gg d$, where $d$ denotes the size of the source and $\vec{\bm{x}}$ is the vector pointing to the observer from the COM, we have  
 \bqn
\lb{3.24a}
\Phi^{\mathrm{IV}}&=&  
\frac{c_{14}-2c_{13}}{2c_{14}(c_{13}-1)}\Phi^{\mathrm{II}},\nb\\
\Psi^{\mathrm{II}}_i&=&-\frac{c_{13}}{1-c_{13}}\Psi^{\mathrm{I}}_i,
 \eqn
and  
 \bqn
 \lb{3.24c}
\Psi^{\mathrm{I}}_{i,j}&=&-\frac{1}{c_V}\dot{\Psi}^{\mathrm{I}}_i N_j,
\nb\\
\Phi^{\mathrm{II}}_{,i}&=&-\frac{1}{c_S}\dot{\Phi}^{\mathrm{II}} N_i,
 \eqn
where $N_k$ denotes the unit vector along the direction between the source (the COM) and the observer, and $\Psi^{\mathrm{I}}_{i}$ is another gauge-invariant quantity defined in \cite{Kai19}
via the relation, 
 \bqn
 \lb{PsiandPhi}
\Psi^{\mathrm{I}}_i \equiv \gamma_i+\nu_i.
 \eqn
Then, inserting the above expressions into   (\ref{3.22}) and
(\ref{3.23})
we obtain
 \bqn
 \lb{3.25}
{{\cal{P}}}_{ij}&=&\phi_{ij}-\frac{2c_{13}}{(1-c_{13})c_{V}}\Psi^{\mathrm{I}}_{(i}N_{j)}\nb\\
&&-\frac{c_{14}-2c_{13}}{c_{14}(c_{13}-1)c^2_{S}}\Phi^{\mathrm{II}}N_iN_j+\delta_{ij}\Phi^{\mathrm{II}}.
 \eqn

 Assuming that  (${\bf e}_X, {\bf e}_Y, {\bf e}_Z$) are three unit vectors that form a set of orthogonal basis with  ${\bf e}_Z \equiv \bf{N}$, so that
 (${\bf e}_X, {\bf e}_Y$) lay on the plane orthogonal to the propagation direction $\bf{N}$ of the gravitational wave, we find that,  in the coordinates $x^{\mu} = (t, x^i)$,
 these three vectors
 can be specified by two angles, $\vartheta$ and $ \varphi$, via the relations \cite{PW14}, 
\bqn
\lb{rotationsA}
{\bf e}_{X} &=& \left(\cos\vartheta\cos\varphi, \cos\vartheta\sin\varphi, - \sin\vartheta\right), \nb\\
{\bf e}_{Y} &=& \left(-\sin\varphi, \cos\varphi, 0\right), \nb\\
{\bf e}_{Z} &=& \left(\sin\vartheta\cos\varphi, \sin\vartheta\sin\varphi, \cos\vartheta\right).
\eqn
Then, we can define the six  GW polarizations $h_{N}$'s by 
\bqn
\lb{polarizations}
h_+&\equiv&\frac{1}{2}\left({\cal{P}}_{XX}-{\cal{P}}_{YY}\right), \quad
h_\times\equiv\frac{1}{2}\left({\cal{P}}_{XY}+{\cal{P}}_{YX}\right), \nb\\
h_b&\equiv& \frac{1}{2}\left({\cal{P}}_{XX}+{\cal{P}}_{YY}\right),  \quad
h_L\equiv {\cal{P}}_{ZZ}, \nb\\
h_X&\equiv&\frac{1}{2}\left({\cal{P}}_{XZ}+{\cal{P}}_{ZX}\right), \quad  
h_Y\equiv \frac{1}{2}\left({\cal{P}}_{YZ}+{\cal{P}}_{ZY}\right),  \nb\\
\eqn
where ${\cal{P}}_{AB} \equiv {\cal{P}}_{ij}e^{i}_A e^{j}_B$, with $A, B = \{X, Y, Z\}$. However,  
 in Einstein-aether theory, only five GW polarizations are independent.  With the help of Eq.~(\ref{3.23}) and some related equations, 
 we find that the above expressions can be written explicitly in the form,
\bqn
\lb{pol2}
&& h_+=\frac{1}{2} \phi_{ij} e_+^{ij},\quad h_\times=\frac{1}{2} \phi_{ij} e_\times^{ij},\nb\\
&& h_b=\frac{1}{2}F,\quad h_L=(1+2 \beta_2) h_b, \nb\\
&& h_{ {X}}=\frac{1}{2} \beta_1 \nu^{i} e_X^{i},\quad h_{ {Y}}=\frac{1}{2} \beta_1 \nu^{i} e_Y^{i},
\eqn
where $e_+^{kl} \equiv e_X^k e_X^l-e_Y^k e_Y^l$ and $e_\times^{kl} \equiv e_X^k e_Y^l+e_Y^k e_X^l$, and
\bqn
\lb{beta1}
 \beta_1 \equiv -\frac{2 c_+}{c_V}, \quad
 \beta_2 \equiv  -\frac{c_{14}-2 c_+}{2 c_{14} (1-c_+) c_S^2}.
\eqn

Observe that these equations for the GW polarizations are quite similar to those found for generic modified gravity theories in Chatziioannou, et al.~\cite{Chat12} (see e.g.~Eq.~8 in~\cite{Chat12}). The main difference here is that Chatziioannou, et al., following Poisson and Will~\cite{PW14}, made the implicit assumption that all GW modes travel at the same speed, and this speed is equal to the speed of light. As we saw in the previous subsection, this is not the case in Einstein-aether theory, with some speeds already stringently constrained but others essentially unconstrained: $ - 3\times 10^{-15} < c_T -1 < 7\times 10^{-16}$  due to GW170817~\cite{Oost18}, which leads to $|c_{13}| = |c_{+}| \lesssim 10^{-15}$, but $c_{V} \sim (c_{1}/c_{14})^{1/2} > 1$ and $c_{S} \sim (c_{2}/c_{14})^{1/2} > 1$ and are essentially unconstrained. Therefore, the results of Chatziioannou, et al.~\cite{Chat12} cannot be straightforwardly applied to Einstein-aether theory, but rather they would have to be extended to allow for modes with different and arbitrary speeds. 
 
 In order to calculate the waveforms, let us first assume that the observers (or detectors) are located in a region far away from the source, $R \equiv |{\vec{x}}| \gg d$, where $d$ is the typical size of the system. Notice that  $R$ used here is not the Ricci scalar used in the previous section, but rather the distance to the source. In this region,  we have a useful mathematical method to solve the wave equations. That is, for equations in the form 
\bqn
\lb{fie4}
 \frac{1}{v_s^2} \ddot{\psi}- \Delta {\psi}&=& 16 \pi \tau, 
\eqn
 where $\psi$, $v_s$ and $\tau$ denote the field we are going to solve for, the speed for the corresponding field, and a source term, respectively, we have the following asymptotic solution \cite{Willb}, 
\bqn
\lb{asy2}
 {{\psi}(t, \vec{\bm{x}})} &=& { \frac{4}{R}} \left[  {\sum_{n=0}^{\infty} \frac{1}{n! v_s^n} \frac{\partial^n}{\partial t^n} \int \tau (t-{R}/{v_s}, |\vec{\bm{x}}^{'}|)}  \right. \nb\\
&& \left.  {\times \left (x^{'i} \cdot \frac{x^i}{R} \right)^n d^3 x^{'} } \right] +  {{\cal{O}} \Big({R}^{-2} \Big).}
\eqn
Then, in the gauge \cite{Foster06},
\bq
\lb{4.13}
\phi_i=0, \quad  \nu=\gamma=0, 
  \eq
 we find that  the wave equations given in the last subsection have the solutions  
  \bqn
\lb{4.10}
\phi_{ij}&=&\frac{2 {G_{\ae}}}{R}(\ddot{Q}_{ij})^{TT},\\
\lb{4.11}
\nu_i&=&-\frac{2 {G_{\ae}}}{(2c_1-c_{13}c_-)R} \left. \bigg[\frac{1}{c_V} \left. {\bigg(\frac{c_{13}}{1-c_{13}} \ddot{Q}_{ij}} \right. \right. 
\nb\\ && \left.\left.  {~~~~~~~~~~~-\ddot{\cal{Q}}_{ij}-{\cal{V}}_{ij} }\right. \bigg)  N^j  {+2 \Sigma^i }\right. \bigg]^T,\nb\\
\gamma_i&=&-c_{13}\nu_i,\\
\lb{4.12}
{F}&=& {\frac{G_{\ae}}{R} \frac{c_{14}}{2-c_{14}}} \left [6 (Z-1) \ddot{Q}_{ij} N^i N^j+2 Z \ddot{I}\right.\nb\\
&&\left.  {~~~~-\frac{4}{c_{14} c_S^2} \ddot{\cal{I}}_{ij} N^i N^j-\frac{8}{c_{14} c_S} \Sigma^i N^i }\right], \nb\\
h_{00}&=&2\omega^0=\frac{1}{c_{14}}F,~~~~\phi=-\frac{1+c_2}{c_{123}}f,
 \eqn
 where 
\bqn
\lb{Qijetc}
 {I_{ij}}&\equiv&  {\sum\limits_A m_A x_A^i x_A^j,} \quad
 {I} \equiv {I_{kk},} \nb\\
 {Q_{ij}}&\equiv&  {I_{ij}-\frac{1}{3} \delta_{ij} I,}\nb\\
 {{\cal{I}}_{ij}} &\equiv&  {\sum\limits_A \sigma_A \tilde{m} x_A^i x_A^j, }\quad
 {\cal{I}} \equiv  {{\cal{I}}_{ii},} \nb\\
 {{\cal{Q}}_{ij} }&\equiv& {{\cal{I}}_{ij}-\frac{1}{3} \delta_{ij} \cal{I},}\nb\\
 {\Sigma^i }&\equiv&  {- \sum\limits_A \sigma_A \tilde{m}_A v_A^i, }\nb\\
 {{\cal{V}}_{ij}} &\equiv&  {2  \sum\limits_A \sigma_A \tilde{m}_A \dot{v}_A^{[i}x_A^{j]}, }
\eqn
 {and}
\bqn
\lb{Z}
 {Z} &\equiv& {\frac{(\alpha_1-2 \alpha_2) (1-c_+)}{3 (2 c_+-c_{14})},}\\
\lb{alpha12}
 {\alpha_1} &\equiv&  {- \frac{8\left(c_1c_{14} - c_{-}c_{13}\right)}{2c_1 - c_{-}c_{13}},} \nb\\
 {\alpha_2} &\equiv&   {\frac{1}{2}\alpha_1  + \frac{(c_{14}- 2c_{13}) (3c_2+c_{13}+c_{14})}{c_{123}(2-c_{14})} {.}}~~~~~
\eqn
  
Finally, we note that for any symmetric tensor $S_{ij}$, we have $S_{ij}^{TT}=\Lambda_{ij,kl}S_{kl}$ and $S_i^{T}=P_{ij}S_j$, where $\Lambda_{ij,kl}$ and $P_{ij}$
are the projection operators   defined, respectively, by  Eqs.~(1.35) and (1.39) in  \cite{Mag08}.

Inserting Eqs.~(\ref{4.10}) - (\ref{4.12}) into (\ref{pol2}) and using the above equations, we find that 
 \bqn
\lb{polarizationsB}
h_+& = &  
 \frac{ {G_{\ae}}}{R}\ddot{Q}_{kl}e_+^{kl},\quad 
h_\times = \frac{ {G_{\ae}}}{R}\ddot{Q}_{kl}e_\times^{kl}, \nb\\
h_b &=&\frac{ {c_{14}G_{\ae}}}{R(2-c_{14})} \Bigg[3(Z-1)\ddot{Q}_{ij}e_Z^ie_Z^j  {+Z\ddot{I}}\nb\\
&&~~~~~~~~~~~~-\frac{4}{c_{14}c_S}\Sigma_i e_Z^i  {-\frac{2}{c_{14} c_S^2} \ddot{\cal{I}}_{ij} N^i N^j} \Bigg], \nb\\
h_L &=& \left[1-\frac{c_{14}-2c_{13}}{c_{14}(c_{13}-1)c_S^2}\right]h_b, \nb\\
h_X &=& {\frac{2c_{13} {G_{\ae}}}{(2c_1-c_{13}c_-)c_V R}}\nb\\
 &&\times \left[{{\frac{e_Z^i}{c_V}} \bigg(\frac{c_{13}}{1-c_{13}} \ddot{Q}_{ij}-\ddot{\cal{Q}}_{ij}-{\cal{V}}_{ij}  \bigg)}-2\Sigma_j\right]e_X^j ,\nb\\
h_Y &=&  \frac{2c_{13}G_{\ae}}{(2c_1-c_{13}c_-)c_V R}\nb\\
 &&\times \left[{{\frac{e_Z^i}{c_V}   } \bigg(\frac{c_{13}}{1-c_{13}} \ddot{Q}_{ij}-\ddot{\cal{Q}}_{ij}-{\cal{V}}_{ij}  \bigg)}-2\Sigma_j\right]e_Y^j .~~~\nb\\
 \eqn

The above expressions differ from the work of Hansen, et al.~\cite{HYY15} because the latter built on the work of Chatziioannou, et al.~\cite{{Chat12}}, 
 which as already explained, cannot be applied to Einstein-aether theory. Note, however, that although some of the dependence of the modes on the
 coupling constants $c_{i}$ are different, the general structure of the solution found by Hansen, et al.~\cite{HYY15} remains correct. For example, as found
 in that paper, and shown again by the above equations, the scalar longitudinal mode $h_L$ is proportional to the scalar mode $h_b$, which then
 means that out of the six possible GW  polarizations, only five are independent. Moreover, as shown again in Hansen, et al.~\cite{HYY15} and also in the equations
 above, the breathing and longitudinal modes are suppressed by a factor $c_{14} \lesssim {\cal{O}}\left(10^{-5}\right)$ {\cite{Oost18}}
with respect to the transverse-traceless modes $h_{+}$ and $h_{\times}$\footnote{The overall $c_{14}$ cancels with $1/c_{14}$ in the last two terms inside the square brackets of $h_b$ in Eq.~\eqref{polarizationsB}. However, $\Sigma_i$ and $\ddot{\cal{I}}_{ij}$ in these terms are proportional to $\sigma \sim s$. The sensitivity $s$ scales with $\alpha_1$ and $\alpha_2$ [see Eq.~\eqref{saA}], which scale with $c_{14}$ when $c_{13} \simeq 0$. }, 
 while the vectorial modes $h_{X}$ and $h_{Y}$ are suppressed by a factor
 $c_{13}\lesssim {\cal{O}}\left(10^{-15}\right)$ \cite{Oost18}.

With the GW polarizations at hand, we can now  move to the calculation of the energy flux. Using the Noether current method described in \cite{Foster07,SY17}, 
we find that the energy loss rate is given by
\bqn
\lb{Edot1}
\dot{E}_{b}&=&-\frac{1}{16 \pi G_{\ae}} \left < \int d\Omega R^2 \left [\frac{1}{2 c_T} \dot{\phi}_{ij} \dot{\phi}_{ij} \right. \right. \nb\\
&& \left. \left. +\frac{(2 c_1-c_{13}  c_-) (1-c_{13} )}{c_V} \dot{\nu}^i \dot{\nu}^i \right.\right.\nb\\
&&\left.\left. + \frac{2-c_{14}}{4 c_S c_{14}}  \dot{F} \dot{F} \right] \right> +\dot{O},
\eqn
where  an overhead dot stands for a time derivative, $\Omega$ is the solid angle, and the {angle} brackets stand for an average over one period, defined by 
\bq
\lb{Average}
\langle{\cal{H}}(t) \rangle \equiv \frac{1}{P_b} \int_0^{P_b} {\cal{H}} (t) dt, 
\eq
with $P_b$ the orbital period \cite{Kramer06}. The last term $\dot{O}$ will be omitted from now on, since its purpose is just to cancel secular terms that arise from the other terms in this equation, as discussed in detail in \cite{Yagi14, Foster07}. Using the mathematical tricks presented  in \cite{Mag08}, we find that Eq.~(\ref{Edot1}) becomes 
\bqn
\lb{Edot2}
\dot{E}_{b}&=&-G_{\ae} \left < \frac{{\cal{A}}_1}{5} \dddot{Q}_{ij} \dddot{Q}_{ij}+ \frac{{\cal{A}}_2}{5} \dddot{Q}_{ij} \dddot{\cal{Q}}_{ij}+ \frac{{\cal{A}}_3}{5} \dddot{\cal{Q}}_{ij} \dddot{\cal{Q}}_{ij} \right. \nb\\
&& \left. +{\cal{B}}_1 \dddot{I} \dddot{I}+{\cal{B}}_2 \dddot{I} \dddot{\cal{I}}+{\cal{B}}_3 \dddot{\cal{I}} \dddot{\cal{I}}+{\cal{C}} \dot{\Sigma}^i \dot{\Sigma}^i +{\cal{D}} \dot{{\cal{V}}}_{ij} \dot{{\cal{V}}}_{ij} \right >,\nb\\
\eqn
where
\bqn
\lb{ABCD}
&&{\cal{A}}_1 \equiv \frac{1}{c_T}+\frac{2 c_{14} c_{13} ^2}{(2 c_1-c_{13}  c_-)^2 c_V}+\frac{3 c_{14} (Z-1)^2}{2 (2-c_{14}) c_S},\nb\\
&&{\cal{A}}_2 \equiv -\frac{2 c_{13} }{(2 c_1-c_{13}  c_-) c_V^3}-\frac{2 (Z-1)}{(2-c_{14}) c_S^3}, \nb\\
&&{\cal{A}}_3 \equiv \frac{1}{2 c_{14} c_V^5}+\frac{2}{3 c_{14} (2-c_{14}) c_S^5}, \nb\\
&&{\cal{B}}_1 \equiv \frac{c_{14} Z^2}{4 (2-c_{14}) c_S}, \nb\\
&&  {\cal{B}}_2 \equiv -\frac{Z}{3 (2-c_{14}) c_S^3}, \nb\\
&&{\cal{B}}_3 \equiv \frac{1}{9 c_{14} (2-c_{14}) c_S^5}, \nb\\
&&{\cal{C}} \equiv \frac{4}{3 c_{14} c_V^3}+\frac{4}{3 c_{14} (2-c_{14}) c_S^3}, \nb\\
&&{\cal{D}} \equiv \frac{1}{6 c_{14} c_V^5}.
\eqn
Note that in the above expressions, we corrected a simple typo (minus signs in $A_{2}$) in previous work~\cite{Yagi14}, which originates from the sign error in \cite{Foster07}, and which has been corrected in Eq.~\eqref{spin11} as already mentioned.  

\subsection{Binary Systems}

In  this subsection, we apply the general formula developed in the last two subsections to a binary system. Before doing so, let us first note that such a problem has already been considered in Hansen, et al. \cite{HYY15}, as discussed earlier. The work in this subsection differs from that of Hansen, et al. in that (i) we include in the calculation of the GW polarization modes the fact that the different fields of Einstein-aether theory travel at different velocities, and (ii) we allow for the COM to not be comoving with the aether, i.e.,~we allow $V^{i} \neq 0$. The latter condition is more general than that adopted previously in the literature, thus allowing for the possibility that the aether flow may be in a different direction as compared to the motion of the COM.   

With the  above in mind, we first assume that the binary components are in a quasi-circular orbit. By  ``quasi-circular" we mean that the two celestial bodies are rotating in a fixed plane and the orbit for its one-body effective model is almost a circle within one period \cite{Marionb}. In addition, we also assume that $\dot{\omega}_s \ll \omega_s^2$, where $\omega_s={2 \pi}/{P_b}$ denotes the orbital angular frequency of the orbit \cite{Mag08}. Then, to leading (Newtonian) order in the PN theory, we have
\begin{align}
\lb{quasi1}
 \dot{v}^i &\equiv \ddot{r}^i \simeq - \frac{{\cal{G}} m}{r^2} \hat{r}^i \left[1 + {\cal{O}}\left(\frac{{\cal{G}} m}{r}\right)\right], \\\
\lb{quasi1-2}
  v^2 &\equiv v^i v^i \simeq \frac{{\cal{G}} m}{r} \left[1 + {\cal{O}}\left(\frac{{\cal{G}} m}{r}\right)\right],
\end{align}
where $r = |x_{1}^{i} - x_{2}^{i}|$ is the distance between the two bodies  and $\hat{r}^i \equiv {r^i}/{r} \equiv ({x_1^i-x_2^i})/{r}$ and $m$ is the total mass. 
Here, the relation between  $\cal{G}$ and $G_N$ is given by
\bqn
\lb{Gsen}
{\cal{G}} \equiv G_N (1-s_1) (1-s_2),
\eqn
where $s_A$ is related to   $\sigma_A$  via the relation, $s_A \equiv \sigma_A/(1+\sigma_A)$. 
In \cite{Yagi14}, the sensitivities for neutron stars were calculated numerically for various choices of the coupling constants $c_i$'s. Unfortunately, all of those choices are out of the currently physically viable region defined in Eq.~(\ref{c1234}). In \cite{Foster07}, an analytical expression in the weak-field approximations was given,
\bqn
\lb{saA}
s_A = \left(\alpha_1 - \frac{2}{3}\alpha_2\right) \frac{\Omega_A}{m_A} + {\cal{O}}\left(\frac{G_N m}{d}\right)^2,
\eqn
where $\Omega_A$ is the binding energy of the $A$-th body \footnote{Note that there is an extra factor $c_{14}$ appearing in Eq. (70) of \cite{Foster07} in the published version, which has been corrected in the arXiv version.} and we recall $d$ represents the characteristic size of the system.
This expression is only valid for weakly-gravitating bodies, and thus, strictly speaking, it does not apply to neutron stars or to black holes when considering strong-field effects; for neutron stars, the sensitivities are about an order of magnitude larger and they depend on the equation of state, while for black holes, they may be identically zero, as is the case in khronometric gravity within a parameter space that is of physical interest~\cite{Ref22}.

Since the choice of coordinates $x^{\mu}$ comoving with the aether [cf. Eq.~(\ref{M4})] 
is fixed only up to the spatial diffeomorphism  ${x'}^i = {x'}^i\left(x^k\right)$, as mentioned earlier, we can use this remaining gauge freedom to choose the spatial coordinates so that the binary system 
is always on the $(x, y)$-plane. This then implies that $\bf{\hat{r}}$ can be parameterized via
\bqn
\lb{rhat}
\mathbf{\hat{r}}=\cos\Phi {\bf{\hat{i}}}+\sin\Phi {\bf{\hat{j}}},
\eqn
where $\Phi (t) \equiv \int^t w_s(t') dt'$ is the orbital phase of the binary system, and ${\bf{\hat{i}}}$, ${\bf{\hat{j}}}$, ${\bf{\hat{k}}}$ are unit vectors along the $x$, $y$ and $z$ directions respectively,
with ${\bf{\hat{k}}}= {\bf{\hat{i}}} \times  {\bf{\hat{j}}}$.

Substituting  the above expressions into Eq.~(\ref{polarizationsB}) and only keeping terms up to relative ${\cal{O}}(v^2)$, where $V^i$ is assumed to be of ${\cal{O}}(v)$, we find,  
\allowdisplaybreaks[4]
\begin{widetext}
\bqn
\lb{hp}
h_+&=&- \frac{2 G_{\ae}}{ R} {\cal{M}} {\cal{U}}^2 (1+\cos^2\vartheta) \cos (2\Theta) 
+ \underbrace{\frac{ 2{G_{\ae}}}{ R} m V^k V^l e_+^{kl}} ,\\
\lb{hc}
h_\times&=&\frac{4 G_{\ae}}{ R} {\cal{M}} {\cal{U}}^2 \cos \vartheta \sin (2\Theta) 
+ \underbrace{\frac{ 2{G_{\ae}}}{ R} m V^k V^l e_\times^{kl}} ,
\\ \lb{hb}
h_b&=&\frac{2 G_{\ae}}{R} \frac{c_{14}}{2-c_{14}}  
\left[ 
\frac{2 \Delta s}{c_{14} c_S} \eta^{1/5} {\cal{M}} {\cal{U}} \sin \vartheta \sin \Theta
\right. \nb\\
&&~~~~~ \left. 
+ \frac{2 {\cal{S}}-3 c_{14} (Z-1) c_S^2}{c_{14} c_S^2} {\cal{M}} {\cal{U}}^2 \sin^2 \vartheta \cos(2 \Theta)
-\frac{4 \Delta s }{c_{14} c_S^2} \eta^{1/5} {\cal{M}} {\cal{U}} (V^i N^i) \sin \vartheta \sin \Theta  \right. \nb\\
&& ~~~~~\left.+ \underbrace{\frac{3 c_{14} c_S^2 (Z-1)-2  {\cal{S}}^\prime}{c_{14} c_S^2} m V^i V^j N^i N^j} 
+ \underbrace{\frac{2  {\cal{S}}^\prime}{c_{14} c_S} m V^i N^i+m V^i V^i} \right], \\
\lb{hL}
h_L&=&\left[1+ \frac{c_{14}-2 c_{13} }{c_{14} (1-c_{13} ) c_S^2} \right] h_b, 
\\
\lb{hx}
h_{ {X}}&=&- \frac{\beta_1 G_{\ae}}{R} \frac{1}{2 c_1-c_{13}  c_-} 
\left[
-2 \Delta s  \eta^{1/5} {\cal{M}} {\cal{U}} \cos \vartheta \sin \Theta 
\right. \nb\\
&& ~~~~~\left. 
+\frac{1}{c_V} \left({\cal{S}}- \frac{c_{13} }{1-c_{13} }\right) {\cal{M}} {\cal{U}}^2 \sin (2 \vartheta) \cos (2\Theta)
-\frac{2 \Delta s}{c_V}  \eta^{1/5} {\cal{M}} {\cal{U}}  \big(\sin \vartheta e_X^i+\cos \vartheta N^i\big) V^i  \sin \Theta 
\right. \nb\\
&& ~~~~~\left. 
-\underbrace{\frac{2 m}{c_V} \left({{\cal{S}}^\prime}- \frac{c_+}{1-c_+} \right) V^i V^j e_X^i N^j} 
  - \underbrace{2{\cal{S}}^\prime m e_X^i V^i} \right],~~~~ 
\\
\lb{hy}
h_{ {Y}}&=&- \frac{\beta_1 G_{\ae}}{R} \frac{1}{2 c_1-c_{13}  c_-} 
\left[
-2 \Delta s  \eta^{1/5} {\cal{M}} {\cal{U}}  \cos \Theta 
\right. \nb\\
&& ~~~~~\left. 
-\frac{2}{c_V} \left({\cal{S}}- \frac{c_{13} }{1-c_{13} }\right) {\cal{M}} {\cal{U}}^2 \sin (\vartheta) \sin (2\Theta)
-\frac{2 \Delta s}{c_V}  \eta^{1/5} {\cal{M}} {\cal{U}} \big(\sin \vartheta \sin \Theta e_Y^i+\cos \Theta N^i\big) V^i 
\right. \nb\\
&& ~~~~~\left. 
-\underbrace{\frac{2 m}{c_V} \left({{\cal{S}}^\prime}- \frac{c_+}{1-c_+} \right) V^i V^j e_Y^i N^j} 
  - \underbrace{2{\cal{S}}^\prime m e_Y^i V^i} \right],~~~~ 
\eqn
\end{widetext}
where
\bqn
\lb{metc}
m \equiv m_1+m_2, \quad \mu_A \equiv \frac{m_A}{m}, \quad \mu \equiv \mu_1 \mu_2 m,\nb\\
\eta \equiv \frac{\mu}{m}, \quad {\cal{M}} \equiv m \eta^{3/5}, \quad {\cal{U}} \equiv ({\cal{G}} {\cal{M}} \omega_s)^{1/3},
\eqn
and
\bqn
\lb{deltas}
&&\Delta s \equiv s_1-s_2, \quad
{\cal{S}} \equiv s_1 \mu_2+s_2 \mu_1, \nb\\
\lb{Theta}
&& \Theta \equiv \varphi-\Phi,  \quad
{{\cal{S}}^\prime \equiv s_1 \mu_1+s_2 \mu_2}.
\eqn

Now several comments are in order. 
First, the above expressions for the plus and cross polarization modes [Eqs.~(\ref{hp}) and (\ref{hc})] reduce to those of GR \footnote{There is a simple transcription typo in~\cite{HYY15}, which accidentally dropped  factor of $\eta^{1/5}$ in these modes.}~\cite{Chat12,HYY15}, when $c_i$'s and $s_i$'s are set to be zero. The quantity $\varphi$ determines the coalescence phase, whose value can be chosen arbitrarily. References~\cite{Chat12,HYY15} use the convention $\varphi=0$, which will be adopted in this paper.
Second, these expressions are also similar to those found in Hansen, et al.~\cite{HYY15} to leading order in the PN expansion. However, since Hansen, et al.~\cite{HYY15} used a formalism that implicitly assumed the speed of all modes is the speed of light, which is not the case in Einstein-aether theory, there are factors of $(c_{T},c_{V},c_{S})$ missing in that work, which we correct here. 
Third, the under-braced terms have not appeared in the literature previously. However,  they will be safely neglected  for our current studies, since they are time-independent, and lead to no contributions to the geodesic deviation equation [Eq.~(\ref{3.22})], as can be seen from Eqs.~(\ref{3.22})-(\ref{3.25}).
Fourth, the above expressions  contain terms that are sub-leading in the PN approximation (i.e.~they are of ${\cal{O}}(v)$ smaller than the leading-order modifications), and these have also never appeared in the literature. This is not just because they are sub-leading in the PN approximation, but also because they depend on the COM velocity $V^i$, which is typically assumed to be of order $10^{-3}$ with respect to the CMB rest frame~\cite{Foster07}, and thus is much smaller than the relative velocity of binary constituents before coalescences. These terms, however, cannot be neglected as they are time-dependent, and proportional to $\cos \Theta,\; \sin \Theta$.
Fifth, strictly speaking, Eqs.~(\ref{hp})-(\ref{hy}) should be evaluated at the retarded time $t_r$, where $t_r \equiv t-R/c_N$, with $c_{N}$ being any of $(c_{T}, c_{V}, c_{S})$, depending on the mode under consideration. 
 
With the above in mind, substituting (\ref{quasi1}), (\ref{quasi1-2}), (\ref{Qijetc}) and (\ref{rhat}) into Eq.~(\ref{Edot2}), we find that
\begin{widetext}
\bqn
\lb{Edot3}
\dot{E}_{b}&=&- \frac{G_{\ae} {\cal{G}}^2 \mu^2 m^2}{r^4} \times \left < \frac{8}{15} ({\cal{A}}_1+{\cal{S}} {\cal{A}}_2+{\cal{S}}^2 {\cal{A}}_3) (12 v^2-11 \dot{r}^2)  +4 ({\cal{B}}_1+{\cal{S}} {\cal{B}}_2+{\cal{S}}^2 {\cal{B}}_3) {\dot{r}}^2  \right. \nb\\
&& ~~~~~ \left. + \frac{1}{5} \Delta s [ 8 ({\cal{A}}_2+2 {\cal{S}} {\cal{A}}_3) (3 \dot{r}^j-2 \dot{r}^i \hat{r}^i \hat{r}^j ) +60 ({\cal{B}}_2+2 {\cal{S}} {\cal{B}}_3)  \dot{r}^i \hat{r}^i \hat{r}^j  ] V^j \right. \nb\\
&& ~~~~~  \left. +\Delta s^2 \left[ \left ( \frac{6}{5} {\cal{A}}_3 +36 {\cal{B}}_3-2 {\cal{D}} \right) (\hat{r}^i V^i)^2+\left ( \frac{18}{5} {\cal{A}}_3+2 {\cal{D}}\right ) V^i V^i  +{\cal{C}} \right]\right >.
\eqn
\end{widetext}
It is interesting to note that this result reduces identically to that found by Yagi, et al~\cite{Yagi14}, since in that work, no assumption was made on the speed of the propagating modes.

Equation~\eqref{Edot3} includes Einstein-\ae ther corrections both at $-1$PN $(\dot{E}_{b} \propto v^8)$ and 0PN $(\dot{E}_{b} \propto v^{10})$ orders. When deriving this equation, we only considered the Newtonian contribution in the conservative sector in Eqs.~\eqref{quasi1} and~\eqref{quasi1-2}. Formerly, the 1PN correction to the conservative dynamics can affect $\dot{E}_{b}$ at 0PN order. This is because such 1PN effect can couple to the $-1$PN dipole radiation in Eq.~\eqref{Edot2} to give rise to a 0PN effect in Eq.~\eqref{Edot3}. We do not include such corrections in this paper since they can never become a dominant correction (as they are 1PN correction to the $-1$PN effect). On the other hand, the 0PN effect included in Eq.~\eqref{Edot3} can dominate the $-1$PN effect when e.g. $s_1 \sim s_2$ and the dipole radiation is suppressed.

\section{Evolution of the Orbital Angular Frequency}
 \renewcommand{\theequation}{4.\arabic{equation}} \setcounter{equation}{0}

The emission of gravitational waves causes the separation of the two bodies in a binary system to shrink, which thus leads  the orbital frequency to grow, until coalescence. In this section, we find the evolution of the orbital angular frequency $\omega_s$ through the use of the energy loss rate.
Note that there is a different, yet equivalent, way to get the same result through the Virial theorem (see, e.g., \cite{Landaub,Mag08}). 

The evaluation of the time-domain waveform requires that one solves the equations of motion in Einstein-aether theory. As explained in the previous section, these equations take on a Newtonian-like form, and their solution can be described effectively by Eq.~\eqref{rhat}. All one needs to prescribe now is the evolution of the orbital angular frequency, which we study here to the leading PN order. This equation can be obtained through the Einstein-aether version of Kepler's law \cite{Mag08},
\bqn
\lb{quasi2}
\omega_s^2&\simeq&\frac{{\cal{G}} m}{r^3},
\eqn
 which yields 
\bqn
\lb{quasi3}
\frac{{\dot{\omega}}_s}{\omega_s}&=&\frac{3}{2} \frac{{\dot{E}}_b}{E_b},
\eqn
where $E_b$ in the denominator is the binding energy~\cite{Yagi14}, namely
\bqn
\lb{Eb}
E_b&=&-\frac{{\cal{G}} \mu m}{2 r}.
\eqn
Substitution of Eqs.~(\ref{Eb}), (\ref{Edot3}), (\ref{quasi2}) and (\ref{rhat}) into Eq.~(\ref{quasi3}) leads to 
\bqn
\lb{omegas1}
({\cal{G}} m)^{2} \dot{\omega}_s &=& ({\cal{G}} m)^2 \frac{d {\omega}_s}{dt} \nb\\
&=& \kappa_1 ({\cal{G}} m \omega_s)^{11/3} \left[1+\epsilon_x  \left( {\cal{G}} m \omega_s\right)^{-2/3} \right]\,, ~~~~~~
\eqn
where
\bqn
\lb{kappa1}
{\kappa}_1 &\equiv&   \frac{48\eta(2 - c_{14})}{5(1-s_{1})(1-s_{2})}\left({\cal{A}}_1+{\cal{S}} {\cal{A}}_2+{\cal{S}}^2 {\cal{A}}_3\right),~~~ \\
\lb{epsilonx}
{\epsilon}_x &\equiv& \frac{\Delta s^2}{ 32 ({\cal{A}}_1+{\cal{S}} {\cal{A}}_2+{\cal{S}}^2 {\cal{A}}_3)} \nb\\
&&\times \left[(21 {\cal{A}}_3+90 {\cal{B}}_3+5 {\cal{D}}) V^i V^i\right.\nb\\
&&\left. -(3 {\cal{A}}_3+90 {\cal{B}}_3-5 {\cal{D}}) (V^3) ^2  
+5 {\cal{C}}\right].
\eqn
We also note that we have used the quasi-circular condition.

Solving Eq.~(\ref{omegas1}) exactly is not possible, but a good approximation to the solution can be obtained when $\epsilon_x$ is small enough, i.e.,~when $\epsilon_{x} \ll 1$. Since ${\cal{A}}_1$ is ${\cal{O}}(1)$ and ${\cal{S}}$, as well as ${\cal{S}}^2$, are suppressed by the sensitivities according to the definition in Eq.(\ref{deltas}), the contribution of the denominator of Eq.~(\ref{epsilonx}) is ${\cal{O}}(1)$. Moreover, by using Eq.~(\ref{ABCD}), we see that the coefficients of the $V^i$-related terms are all of ${\cal{O}}(c_{14}^{-1} c_{V, S}^{-5})$, while  ${\cal{C}}$  is of ${\cal{O}}(c_{14}^{-1} c_{V}^{-3}+c_{14}^{-1} c_S^{-3})$. Now recall that for $|c_{13}| \lesssim 10^{-15}$ we  have $c_S \simeq {\cal{O}}(c_2/c_{14})^{1/2}$ and $c_V \simeq {\cal{O}}(c_1/c_{14})^{1/2}$, as one can see from Eqs.~(\ref{cV}) and (\ref{cS}). Thus, because $V^i$ is assumed to be of ${\cal{O}}(v)$ or smaller (see \cite{Foster07}), the contribution from the numerator is of ${\cal{O}}(\Delta s^2 {\cal{C}})$. Putting everything together and using the expressions for ${\cal{C}}$, we first find that
\begin{equation}
\epsilon_{x} \leq  \frac{5}{24} \Delta s^2 c_{14}^{1/2}  \left(c_1^{-3/2}+c_2^{-3/2}\right).
\end{equation}
Observe that if $\Delta s^{2} \ll 1$, either because $s_{1} = 0 = s_{2}$ (as may be the case in black hole binaries) or because $s_{1} = s_{2}$ (equal-mass neutron star binaries), then $\epsilon_{x}$ is always small and the approximation is automatically well-justified. Moreover, if we insert the weak-field limit for the sensitivities in Eq.~(\ref{saA}), the above expression could be further written as
\begin{eqnarray}
\epsilon_{x} &\leq&  \frac{605}{216} c_{14}^{5/2} \left(\frac{\Omega_1}{m_1}-\frac{\Omega_2}{m_2}\right)^2 \left(c_1^{-3/2}+c_2^{-3/2}\right) \nb \\
&\leq& 7 \times 10^{-5},
\end{eqnarray}
where we have used that $c_{14} \lesssim 2.5 \times 10^{-5}$ and $c_{1, 2} \gtrsim c_{14}$ from Eq.~\eqref{c1234}, and that $\Omega_{A} \leq m_{A}$. Clearly then, the above analysis justifies the search for a perturbative solution to Eq.~\eqref{omegas1} in $\epsilon_{x} \ll 1$. 

Even though the requirement that $\epsilon_{x} \ll 1$ is satisfied when one saturates current constraints on the theory, a perturbative solution to Eq.~(\ref{omegas1}) actually requires 
\bqn
\lb{epsilonx2}
 \epsilon_{x} \ll ({\cal{G}} m \omega_{s})^{2/3},
\eqn
which may be more severe when the binary's orbital velocity is small enough. Notice, however, that this implies that $v \gtrsim 0.05$, which is true in the regime of interest of the second-generation ground-based gravitational wave detectors. In such a region, we can perturbatively expand the solution to find 
\begin{align}
\lb{omegas2}
\omega_s (t) &\simeq \kappa_2^{-3/8} ({\cal{G}} m)^{-5/8}  (t_c-t)^{-3/8}
\nb \\
&\times \left [1-\frac{3}{10} \epsilon_x \kappa_2^{1/4} \left(\frac{t_c-t}{{\cal{G}} m}\right)^{1/4} \right], 
\end{align}
where
\bqn
\lb{kappa2}
{\kappa}_2 &\equiv&  \frac{128\eta(2 - c_{14})}{5(1-s_{1})(1-s_{2})}  \left({\cal{A}}_1+{\cal{S}} {\cal{A}}_2+{\cal{S}}^2 {\cal{A}}_3\right),~~~~~~
\eqn
and $t_c$ is the time of coalescence. Clearly, the above results reduce to the well-known expression~\cite{Mag08},
\bqn
\lb{omegasGR}
\omega_s^{GR} (t) &=& \frac{1}{8} \left(\frac{\eta}{5}\right)^{-3/8} (G_{N} m)^{-5/8} (t_c^{GR}-t)^{-3/8},~~~~~~
\eqn 
in the GR limit.

\begin{figure}[htb]
\includegraphics[width=\linewidth,clip=true]{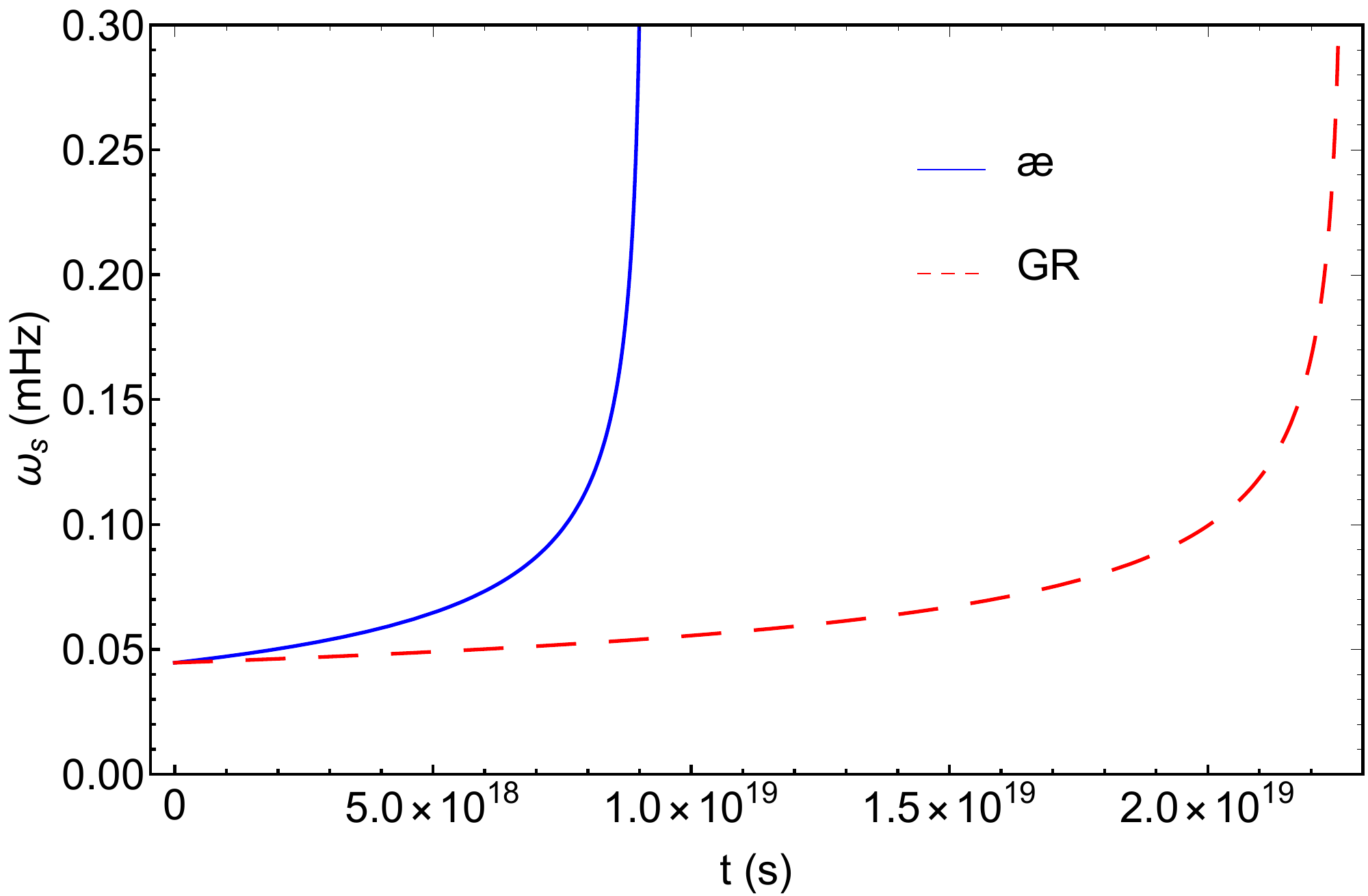} 
\caption{Evolution of the orbital angular frequency $\omega_s(t)$ of the inner binary in the hierarchy triple system J0337 starting at 01-04-2012 to the binary's final stage~\cite{Ransom14}, as given
by Eqs.~(\ref{omegas2}) and (\ref{omegasGR}) for $\ae$-theory and GR, respectively.  It is clear that the orbital angular frequency grows and becomes unbounded at  the coalescence time. Note, however, that the coalescence time for the two theories is different ($t_{c} \sim  9 \times 10^{18}$ and $t_{c}^{GR} \sim 2 \times 10^{19}$ s), because the additional polarization modes of Einstein-aether theory cause the binary to lose binding energy faster than in GR, thus forcing the binary to merge earlier.
 } 
\label{pws}
\end{figure}

Fig.~\ref{pws} shows the difference between the GR and $\ae$-theory evolution of the orbital angular frequency \footnote{In plotting Fig. \ref{pws}  we just used the time coordinate $t$, instead of the retarded time,  $t_{r}^A \equiv t- {R}/{c_A}$ \cite{Jackb}. Since  $\omega_s$ is a function of $(t_c-t)$, there is no difference, as $t_c-t=(t_c- {R}/{c_A})-(t- {R}/{c_A})=t_{r,c}^A-t_r^A$  {(the subscript ``A" here is to distinguish the different kinds of propagation modes: scalar, vector and tensor).}} for the inner binary in the hierarchy triple system PSR J0337+1715 (denoted J0337 henceforth)~\cite{Shao, Ransom14}. Specifically, we set $m_1=1.4378 M_{\odot}$,  $m_2=0.19751 M_{\odot}$, and $\omega_s (t=0) \approx 0.0000446$ Hz {, where $t=0$ stands for the time that J0337 was first observed}. Moreover, we choose the coupling constants to be $c_1=4\times 10^{-5}$, $c_2=9\times 10^{-5}$, $c_4=-2\times 10^{-5}$ and $c_3=-c_1$ as in~\cite{Kai19}, which satisfies all constraints \cite{Oost18}. For the COM velocity, we choose $\vec{V} = (0.002, 0.01, 0.03)$, which satisfies the constraints given  in \cite{Foster07}. The sensitivities of neutron stars are not known in this region of parameter space, so for illustrative purposes only, we use there the weak-field expression of Eq.~\eqref{saA}, with $\Omega_{A}/m_{A} = G_{N} m_{A}/R_{A}$ and $(R_{1}, R_{2}) = (12.7, 6.33\times10^4)$ km. These parameter choices satisfy the perturbative condition $\epsilon_{x} ( {\cal{G}} m \omega_{s})^{-2/3} \ll 1$ for about 1/1000 of its life time, i.e., the duration from the date J0337 was first observed in 2012 to its future merger. Because the time to merger is so long, the parameter choices satisfy the perturbative condition $\epsilon_{x} ( {\cal{G}} m \omega_{s})^{-2/3} \ll 1$ during a time much longer than the designed observing window of LISA-like detectors.

Once $\omega_s$ is known, one can insert it into Eqs.~(\ref{hp}) - (\ref{hy}) to find the GW polarizations. Given the large number of cycles present in these time-domain waveforms, however, it is impractical to plot them straight as functions of time. A better alternative is to decompose the signals into an amplitude and a phase, via
\bqn
\lb{amplitude1}
h_+ &\equiv& A_+ \cos (2 \Theta), \nb\\
h_\times &\equiv& A_\times \sin (2 \Theta), \nb\\
h_b &\equiv& A_{b2} \cos (2 \Theta)+ A_{b1} \sin (\Theta), \nb\\
h_L &\equiv& A_{L2} \cos (2 \Theta)+ A_{L1} \sin (\Theta).
\eqn
Recall that the phase $\Theta$ here is defined from the orbital phase $\Phi$ through Eq.~(\ref{Theta}). Figures~\ref{fig-amp} and~\ref{fig-phase} show the time-domain amplitudes and orbital phase for a binary with the same parameters as those chosen in Fig.~\ref{pws}. In addition, we have here chosen $\vartheta = 39.254$ degrees, according to~\cite{Ransom14}, and $\varphi = 70$ degrees as an illustrative example. To more clearly see the difference between the GR and the $\ae$-theory evolution, we also plot the amplitudes in the GR limit (see also Eq.~(4.29) of \cite{Mag08}). 

\begin{figure*}[htb]
\includegraphics[width=\columnwidth]{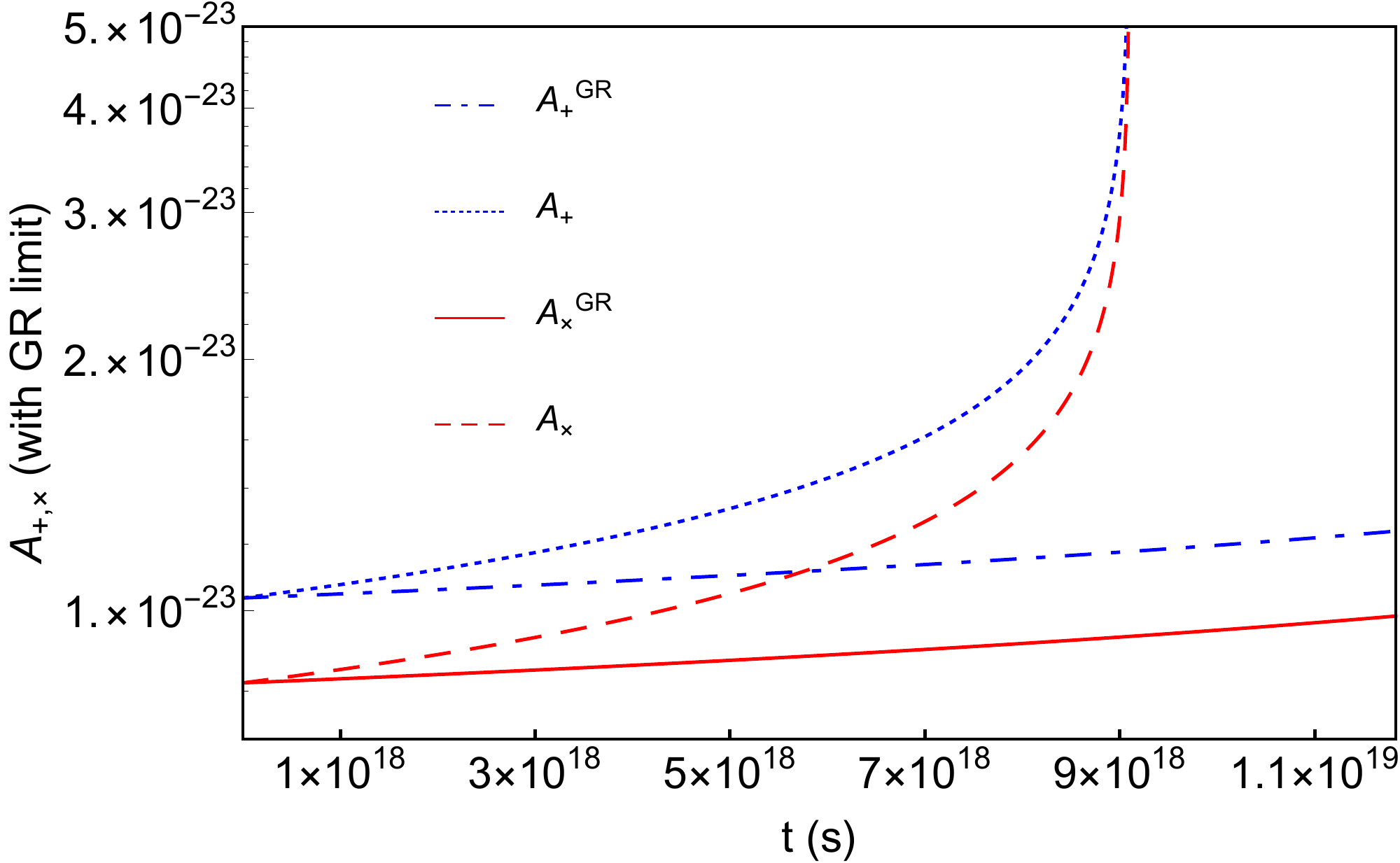} 
\includegraphics[width=\columnwidth]{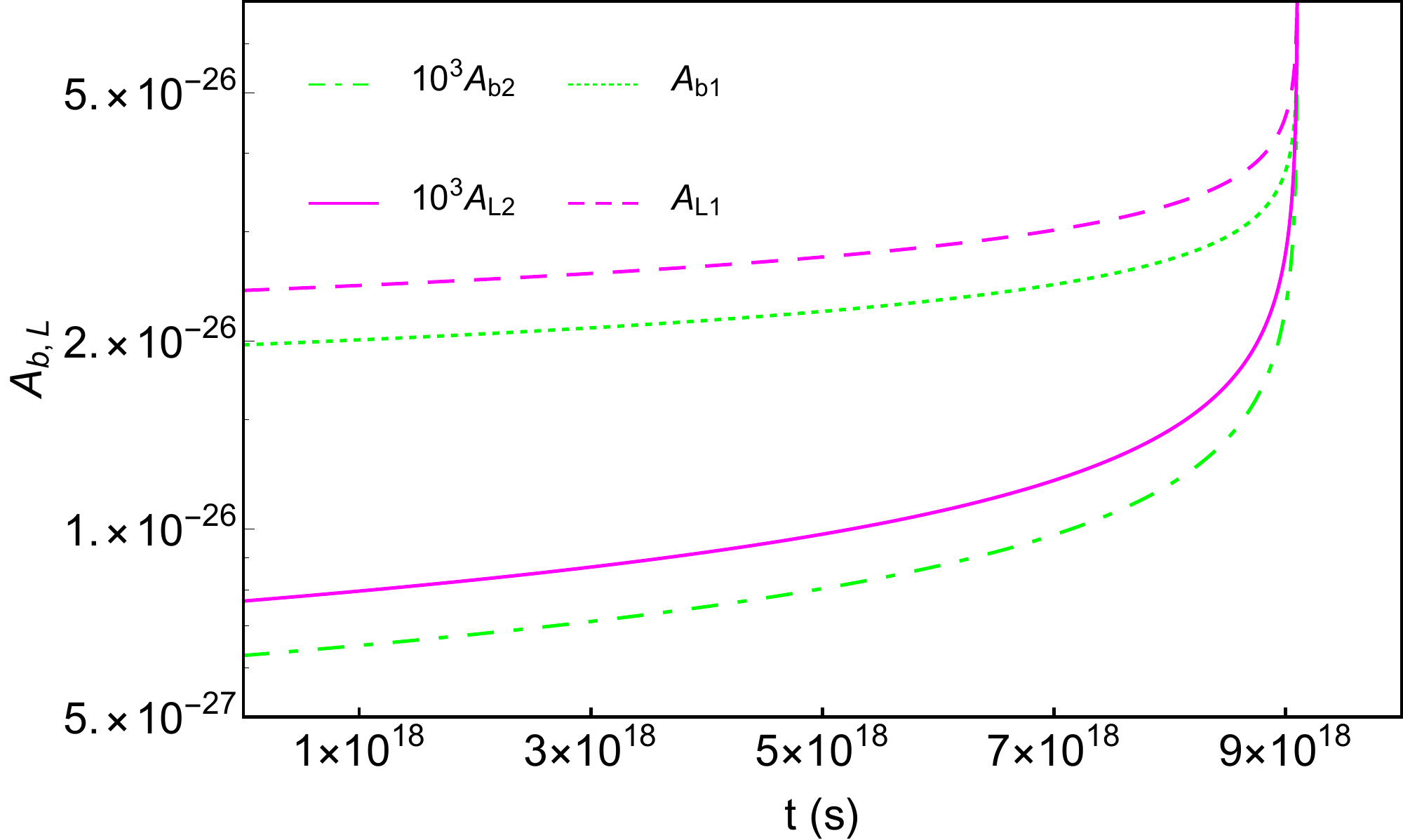} 
\caption{Temporal evolution of the amplitudes of the GW polarizations for the inner binary in the hierarchy triple system J0337~\cite{Ransom14}. The left panel shows the $+$ and $\times$ modes in GR and in $\ae$-theory.  The right panel shows the breathing and longitudinal modes in $\ae$-theory, where the subscript 1 and 2 correspond to the harmonic number. Observe that the second harmonic is rescaled by a factor of $10^3$ relative to the first harmonic, which implies the latter is much larger. Observe also that the amplitudes in $\ae$-theory diverge faster than in GR because the binary inspirals more rapidly.
}
\label{fig-amp}
\end{figure*}

\begin{figure}[htb]
\includegraphics[width=\linewidth,clip=true]{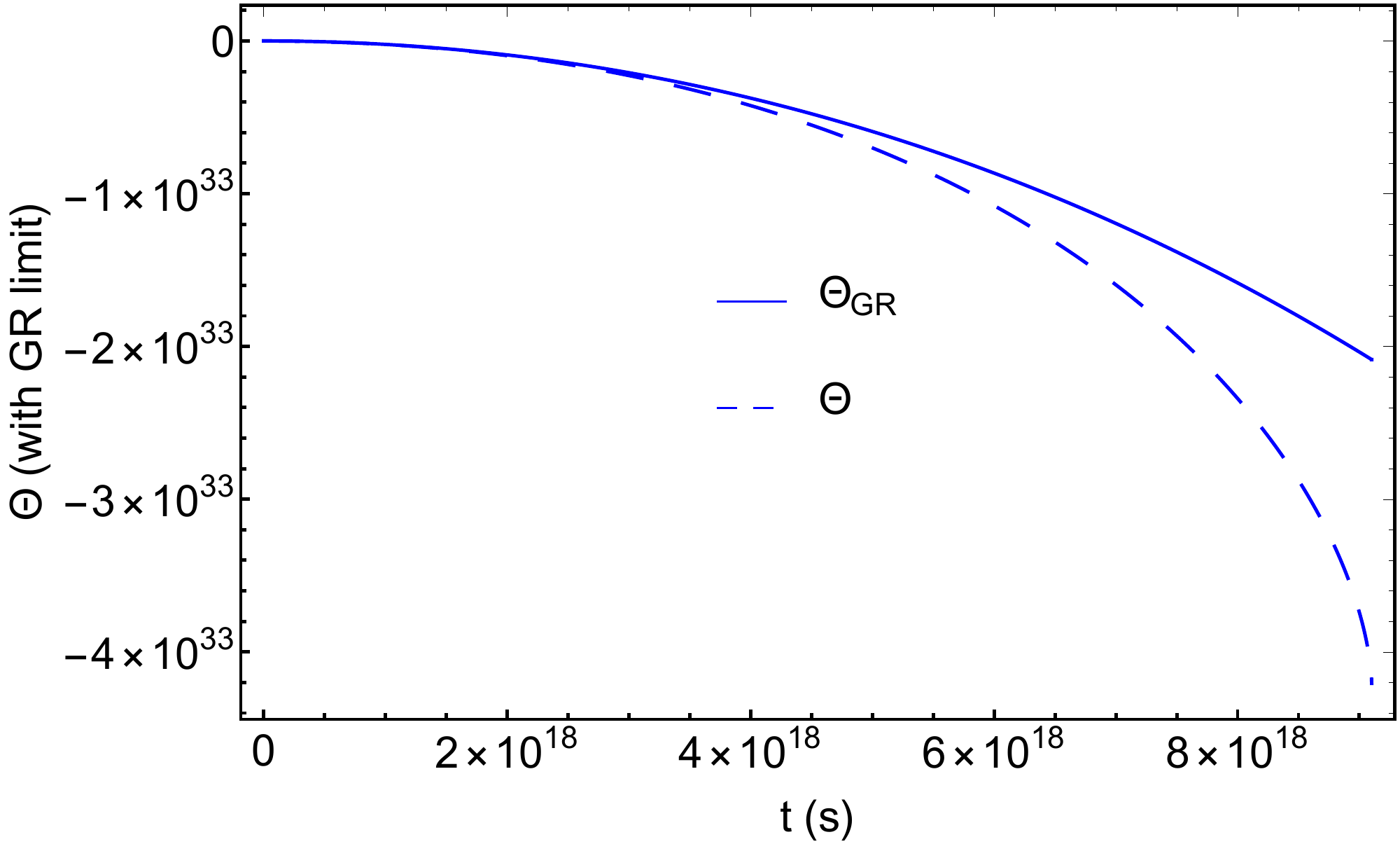} 
\caption{Temporal evolution of the phases of the GW polarizations for the inner binary in the hierarchy triple system J0337 \cite{Ransom14}  in GR and in $\ae$-theory. Note that the phases here are different from the orbital phases in Eq.~(\ref{Theta}), although the differences are trivial. 
 } 
\label{fig-phase}
\end{figure}

These figures deserve several comments. First, notice that with the choice of parameters we have made to make these figures (specifically with $c_{13} = 0$), the $h_{X,Y}$ modes vanish identically. Even if we had saturated current constraints by setting $c_{13} = 10^{-15}$, the amplitudes of these vector modes would be suppressed by at least 15 orders of magnitude relative to the plus and cross modes. The implication then is clear: GW interferometers will never be able to detect these modes directly. 
Second, observe that the scalar modes $h_{b,L}$ are suppressed relative to the tensor modes $h_{+, \times}$ by a factor of $10^{3}$. This then implies that it will be extremely difficult for GW detectors to measure these modes directly. However, we observe from the figures that the amplitude and the phase of the tensor mode is clearly modified, and this is a feature that could be constrained with GW instruments. This is true especially for BNSs since the approximation in Eq.~(\ref{saA}) is better in that case, as discussed previously. Therefore, in the case of Einstein-aether theory, it is clear that constraints on the temporal (or frequency) evolution of the tensor modes are much more constraining than any polarization test that proves that GW signals only contain $+$ and $\times$ modes.


\section{Response function}
 \renewcommand{\theequation}{5.\arabic{equation}} \setcounter{equation}{0}
 
 Gravitational waves emitted by massive binary systems have attracted a lot of attention recently, as they could be ideal sources for both ground- and space-based detectors, such as LIGO, Virgo, KAGRA, LISA, TianQin, Taiji and DECIGO \cite{AS16}. Therefore, in this section we consider the response function for both kinds of detectors.

\subsection{Ground-based L-Shape Detectors}

With the expressions for the GW polarization modes in the coordinate space in hand, we are ready to calculate  the response function $h(t)$ and its Fourier transform  $\tilde{h}(f)$. In this subsection, 
we shall focus on L-shape detectors,  such as LIGO, Virgo and KAGRA \cite{Akutsu}.  From \cite{PW14, Chat12}, we find
\bqn
\lb{RF1}
h(t) &=& \sum_{N}{F_N(\theta, \phi, \psi)  \; h_N(t)},  
\eqn
where
\bqn
\lb{RF2}
&&F_+ \equiv \frac{1}{2} (1+\cos^2 \theta) \cos 2 \phi \cos 2 \psi-\cos \theta \sin 2 \phi \sin 2 \psi, \nb\\
&&F_\times \equiv \frac{1}{2} (1+\cos^2 \theta) \cos 2 \phi \sin 2 \psi+\cos \theta \sin 2 \phi \cos 2 \psi, \nb\\
&&F_b \equiv -\frac{1}{2} \sin^2 \theta \cos 2 \phi, \quad F_L \equiv \frac{1}{2} \sin^2 \theta \cos 2 \phi, \nb\\
&&F_{ {X}} \equiv - \sin \theta (\cos \theta \cos 2 \phi \cos \psi-\sin 2 \phi \sin \psi), \nb\\
&&F_{ {Y}} \equiv - \sin \theta (\cos \theta  \cos 2 \phi \sin \psi+\sin 2 \phi \cos \psi).
\eqn
Here $\{ \theta, \phi, \psi \}$ are the three angles (polar, azimuthal and polarization angles) that specify the relative orientations of the detector with respect to the source [note that the angle $\phi$ here is not the same as the metric perturbation $\phi$ used in Eq.~(\ref{per2})].  Their definitions can be found in \cite{PW14} (see, for example, Fig. 11.5 in that reference).
To calculate   the Fourier transform (FT) of the response function $h(t)$, we shall adopt the  SPA \cite{Chat12,YBW09,Tan18}. In Appendix A, we present a brief summary 
of this method. For more  details, we refer readers to  \cite{Chat12,YBW09,Tan18} and references therein. 

Let us first write Eq.~(\ref{RF1}) in the form, 
\bqn
\lb{RF4}
h(t)   &\equiv& \sum_{N}{H_N(t)}, 
\eqn
where $H_N(t) \equiv F_N h_N(t)$, and $N$ ranges over all the polarization modes, i.e., $N \in (+,\times,b,L,X,Y)$. We can then define the Fourier transform $\tilde{h}(f)$ as
\bqn
\lb{FT1}
&&\tilde{h}(f)   \equiv \int h(t) e^{i 2 \pi f t} dt = \sum_{N}{\tilde{H}_N(f)}, 
\eqn
where $\tilde{H}_N(f)$ is the Fourier transform of $H_N(t)$. Note that the above definition is slightly different from the one used in \cite{Kai19,Zhao19}. For computational convenience, let us also rewrite $H_N(t)$ as
\bqn
\lb{RF5}
H_{N}(t) &=& \left[q_{N(1)}  \cos(2 \Phi)  + q_{N (2)} \sin(2 \Phi)\right]\omega_s^{2/3} \nb\\
&& +\left[q_{N (3)}  \cos\Phi   + q_{N (4)}   \sin \Phi\right] \omega_s^{1/3},
\eqn
where $\omega_s$ and $\Phi$ are all functions of time, and  $q_{N (n)}$ are time-independent \footnote{For detectors, such as  LIGO, Virgo and KAGRA, one can treat $q_{N (l)}$ as time-independent, since their observation windows are very short \cite{Berti18}. However, for detectors like LISA, this approximation needs to be relaxed, as we will discuss in the next subsection.}, and given explicitly in Appendix B.

To apply the SPA to our problem, we need to find $t$ and $\dot{\omega}_s$ as functions of $\omega_s$. Inverting Eq.~\eqref{omegas2} perturbatively in $\epsilon_x \ll 1$, we find
\begin{align}
\lb{omegas3}
t - t_c &= -\frac{3}{8} \frac{1}{\kappa_{1} \omega_s} ({\cal{G}} m \omega_s)^{-5/3} 
\\ \nb
&\times  \left [1 - \frac{4}{5} \epsilon_x  ({\cal{G}} m \omega_s)^{-2/3} + {\cal{O}}[({\cal{G}} m \omega_s)^{2/3},\epsilon_{x}] \right].
\end{align}
But note very importantly that the time-domain waveform is to be evaluated at retarded time, thus $t \to t - R/c_{N}$ when evaluating the orbital phase in the integrand of the Fourier integral. Typically, the factor of $R/c_{N}$ is re-absorbed in the time of coalescence $t_{c}$ because it is a constant, but in $\ae$-theory, this constant will be {\it{different}} for each of the modes present in the response function, and thus, more care must be taken. 

With the results given in Eqs.~(\ref{omegas1}) and (\ref{omegas3}), we are now able to apply the SPA to Eq.~(\ref{RF5}) by following the procedure outlined in Appendix A. After simple but tedious calculations, we find   
\begin{widetext}
\begin{align}
\label{eq:h-FT-SPA}
\tilde{h}(f) = \sum_{N} & \Bigg\{ \frac{\sqrt{\pi}}{2} \left({\cal{G}} m\right)^{1/3} \kappa_1^{-1/2} (q_{N (1)}+i q_{N (2)})  ({\cal{G}} \pi m f)^{-7/6}    \left [1-\frac{1}{2} ({\cal{G}} \pi m f )^{-2/3} \epsilon_x \right ] e^{-i 2 \pi f R \left(1-c_N^{-1}\right)} e^{i \Psi_{(2)}} ,  \nb\\
 +  &  \frac{\sqrt{\pi}}{4} \left({\cal{G}} m\right)^{2/3} \kappa_1^{-1/2} (q_{N (3)}+i q_{N (4)}) ({\cal{G}} \pi m f)^{-3/2}    \left [1-\frac{1}{2} (2 {\cal{G}} \pi m f)^{-2/3} \epsilon_x \right ]  e^{-i 2 \pi f R \left(1-c_N^{-1}\right) e^{i \Psi_{(1)}} } \Bigg\}, 
\end{align}
\end{widetext}
where $N \in (+,\times,b,L,X,Y)$ and $c_+=c_{\times}=c_T$,  $c_b=c_L=c_S$,  $c_X=c_{Y}=c_V$ \footnote{Note that the $c_+$ here is the speed of plus mode instead of the constant $c_{13}$ as in \eqref{cT}.}. The $e^{-i 2 \pi f R \left(1-c_N^{-1}\right)}$ term exists because of the retarded time argument discussed above (see also Appendix A for a more detailed discussion). The Fourier phases $\Psi_{(1)}$ and $\Psi_{(2)}$, corresponding to the first and second harmonics of the orbital period respectively, are given by 
\begin{widetext}
\begin{align}
\lb{Psi12}
\Psi_{(2)}  &\equiv \frac{9}{20} \kappa_1^{-1} ( {\cal{G}} \pi m f)^{-5/3} \left[1-\frac{4}{7}({\cal{G}} \pi  m f)^{-2/3} \epsilon_x 
\right]  +  2 \pi f \bar{t}_c -2 \Phi(t_c)-\frac{\pi}{4},
\nb \\
\Psi_{(1)} &\equiv \frac{9}{40} \kappa_1^{-1} (2 {\cal{G}} \pi m f)^{-5/3}  \left[1-\frac{4}{7}(2 {\cal{G}} \pi m f)^{-2/3} \epsilon_x \right] + 2 \pi f \bar{t}_c -\Phi(t_c)-\frac{\pi}{4}, 
\end{align}
\end{widetext} 
where we have redefined the coalescence time via $\bar{t}_c \equiv t_c+R$.

 Note that the above expressions are different from the ones given in Eqs.~(66) - (74) in \cite{HYY15} because here we do not assume the different polarization modes travel all at the speed of light. Moreover, in our calculation of the Fourier amplitudes, we have included Einstein-aether corrections of ${\cal{O}}(v)$ relative to the leading-order correction. Therefore, while in~\cite{HYY15} the non-tensor modes are all proportional to the first harmonic, here we also have contributions that are proportional to the second harmonic, i.e.~$q_{b,L,X,Y (1)} \neq 0 \neq q_{b,L,X,Y (2)}$. Finally, Eq.~(\ref{eq:h-FT-SPA}) contains a term proportional to $\exp[-2 \pi i f R(1-1/c_{N})]$, which was absent from previous studies because all modes were assumed to travel at the speed of light.

 {We would also like to note that in the present case since now the breathing and longitudinal modes are degenerate [cf. Eq.\eqref{pol2}], the $q_{b(i)}$ and $q_{L(i)}$ terms in Eq.\eqref{eq:h-FT-SPA} can be combined together to simplify the results,}
\bqn
\lb{qS}
&&  {q_{S(i)} \equiv q_{b(i)}+q_{L(i)}=q_{b(i)} (1-a_{bL}), }
\eqn
 {where Eqs.\eqref{qq}, \eqref{ddb} and \eqref{RF2} had been used and $a_{bL}$ is given by \eqref{abL}.} 

\subsection{Space-Based Equilateral-Shape Detectors}

In this subsection we calculate the response function for a space-based equilateral-shape detector, such as LISA, TianQin, Taiji and DECIGO \cite{Kremer,Hu18,Guo18,Seto:2001qf}. Because all such detectors share many similarities in their construction, we will mainly focus on calculations for LISA; similar work applicable to TianQin can be found in~\cite{Hu18, Gong19} for GR. 

Following \cite{Cut98}, we can cast the response function of LISA in the following form, which is similar to Eq.~(\ref{RF4}),
\bqn
\lb{RFL1}
h^{\prime}(t) &=& \frac{\sqrt{3}}{2} \sum_{N} H^{\prime}_N(t),
\eqn
where $N \in (+,\times,b,L,X,Y)$, and where $H^{\prime}_N(t)$  is given by
\bqn
\lb{RFL2}
H^{\prime}_{N}(t) &=& \left[q^{\prime}_{N(1)}  \cos(2 \Phi+\Phi_{DN(2)}) \right. \nb\\
&&+ \left. q^{\prime}_{N (2)}  \sin(2 \Phi+\Phi_{DN(2)}) \right] \omega_s^{2/3}  \nb\\
&&+ \left[q^{\prime}_{N (3)}  \cos(\Phi+\Phi_{DN(1)}) \right. \nb\\
&&+ \left. q^{\prime}_{N (4)} \sin(\Phi+\Phi_{DN(1)})\right]  \omega_s^{1/3},  
\eqn
and the $q^{\prime}_{N(l)}$ expressions are explicitly given in Appendix C. Note that the latter are now functions of time, unlike for ground-based L-shape detectors, as mentioned previously. This is due to the fact that the observational windows of LISA is relatively long and sometimes comparable to the orbital period of the detector. 

The quantities $\Phi_{DN(2)}$ and $\Phi_{DN(1)}$ are the corresponding Doppler phases due to the motion of the detector around the Sun;  gravitational waves reach LISA and the Solar System barycenter at different times \cite{Cut98}. Using the geometry of LISA, we can show that to first-order of $r_{so}/\lambda_N$, where $\lambda_N$ is the wavelength of the $N$-th mode \cite{Cut98} and $r_{so}$ is the radius of the center of mass of LISA which is equal to 1AU, we have \footnote{For the basic construction of LISA, readers are referred to Figs. 1 and 2 of \cite{Cut98}.}, 
\bqn
\lb{Doppler}
\Phi_{DN(2)} &=& \frac{2\omega_{s}}{c_N} r_{so} \sin \bar \theta \cos[\bar \Phi (t)-\bar \phi],\nb\\
\Phi_{DN(1)} &=& \frac{\omega_{s}}{c_N} r_{so} \sin \bar \theta \cos[\bar \Phi (t)-\bar \phi].
\eqn
~~

The quantities $\bar \theta$ and $\bar \phi$ are generated in the same way as in Fig. 11.5 of \cite{PW14} (see also \cite{Cut98}). The quantity $\bar \Phi (t)$ is the orbital phase of the center of mass of LISA in its orbit around the Sun, which is given by
\bqn
\lb{Phibar}
\bar \Phi (t) &=& \bar \Phi_0+\frac{2 \pi t}{T_0},
\eqn
where $\bar \Phi_0$ is a constant and $T_0$ is the period of LISA around the Sun, which is equal to the sidereal period of Earth \cite{Carroll}. 

Since detector-related quantities should be evaluated at the current time $t$, and source-related quantities should be evaluated at the retarded time, one finds that  Eq.~(\ref{RFL2}) needs to be modified to
\bqn
\lb{RFL3}
H^{\prime}_{N}(t) &=& Q_{N(1)}|_{t} \cdot  [\omega_s^{2/3} \cos(2 \Phi)]|_{t_{rN}} \nb\\
&& + Q_{N(2)}|_{t} \cdot  [\omega_s^{2/3} \sin(2 \Phi)]|_{t_{rN}} \nb\\
&& +Q_{N(3)}|_{t} \cdot  [\omega_s^{1/3} \cos \Phi]|_{t_{rN}} \nb\\
&& + Q_{N(4)}|_{t} \cdot  [\omega_s^{1/3} \sin \Phi]|_{t_{rN}},
\eqn
where
\bqn
\lb{QQ}
Q_{N(1)} &\equiv& [q^{\prime}_{N(1)} \cos \Phi_{DN(2)}+q^{\prime}_{N(2)} \sin \Phi_{DN(2)}], \nb\\
Q_{N(2)} &\equiv& -[q^{\prime}_{N(1)} \sin \Phi_{DN(2)}-q^{\prime}_{N(2)} \cos \Phi_{DN(2)}], \nb\\
Q_{N(3)} &\equiv& [q^{\prime}_{N(3)} \cos \Phi_{DN(1)}+q^{\prime}_{N(4)} \sin \Phi_{DN(1)}], \nb\\
Q_{N(4)} &\equiv& -[q^{\prime}_{N(3)} \sin \Phi_{DN(1)}-q^{\prime}_{N(4)} \cos \Phi_{DN(1)}],  \nb\\
\eqn
with $t_{rN} \equiv t - R/c_{N}$.

With the above expressions, we are now in the position to calculate the Fourier transform of LISA's response function using Eq.~(\ref{FT1}) and the SPA technique introduced in Appendix A. The final result is 

\begin{widetext}
\begin{align}
\lb{hfprime}
 {\tilde{h}^\prime(f) }=\frac{\sqrt{3}}{2} \sum_{N}  & \left\{ \frac{\sqrt{\pi}}{2} \left({\cal{G}} m\right)^{1/3} \kappa_1^{-1/2}  \left[Q_{N(1)} |_{t_{a2}+R/c_{N}}+i Q_{N(2)}|_{t_{a2}+R/c_{N}}\right]  ({\cal{G}}\pi m f)^{-7/6}  \right. \nb\\
& \left. \times \left [1-\frac{1}{2} ({\cal{G}} \pi m f )^{-2/3} \epsilon_x \right ]e^{-i 2 \pi f R \left(1-c_N^{-1}\right)} e^{i \Psi_{(2)}} \right. \nb\\
+ & \left. \frac{\sqrt{\pi}}{4} \left({\cal{G}} m\right)^{2/3} \kappa_1^{-1/2} \left[Q_{N(3)} |_{t_{a1}+R/c_{N}}+i Q_{N(4)}|_{t_{a1}+R/c_{N}}\right] ({\cal{G}} \pi mf)^{-3/2} \right. \nb\\
& \left. \times  \left [1-\frac{1}{2} (2 {\cal{G}} \pi m f)^{-2/3} \epsilon_x \right ]  e^{-i 2 \pi f R \left(1-c_N^{-1}\right)} e^{i \Psi_{(1)}} \right\}, 
\end{align}
\end{widetext}
where again $N  \in (+,\times,b,L,X,Y)$, the $\Psi_{(i)}$ are given by Eq.~(\ref{Psi12}), and $t_{a1}$ and $t_{a2}$ are the stationary points {[cf. Appendix A]}. From Eqs.~(\ref{omegas3}) and  (\ref{omegas1}) we find
\begin{align}
\lb{ta12}
t_{a1,2} - t_{c} &= -\frac{3}{8} \frac{1}{\kappa_1 \omega_s(t_{a1,2})} \left({\cal{G}} m \omega_s(t_{a1,2})\right)^{-5/3} 
 \nb\\
& \times \left [ 1 - \frac{4}{5} \epsilon_x \left({\cal{G}} m \omega_s(t_{a1,2})\right)^{-2/3}  \right]\,,  
\end{align}
where $\omega_s(t_{a2}) = \pi f$ and $\omega_s(t_{a1}) = 2 \pi f$.

 {Just like in Eq.\eqref{eq:h-FT-SPA}, the $Q_{b(i)}$ and $Q_{L(i)}$ terms in Eq.\eqref{hfprime} could be combined together, too, since the breathing and longitudinal modes are degenerate [cf. Eq.\eqref{pol2}],}
\bqn
\lb{QS}
&&  {Q_{S(i)} \equiv Q_{b(i)}+Q_{L(i)}=Q_{b(i)} (1-a_{bL}), }
\eqn
 {where Eqs.\eqref{qqL}, \eqref{ddb} and \eqref{RF2} have been used and $a_{bL}$ is given by \eqref{abL}.}

\section{{{Parameterized Post-Einsteinian  Parameters}}}
 \renewcommand{\theequation}{6.\arabic{equation}} \setcounter{equation}{0}
 
 By using the results given in the previous section, we are ready to calculate the ppE  parameters of $\ae$-theory~\cite{YP09, HYY15, Chat12}. Since the calculations for LISA-like detectors are too complicated, we will just focus here on the ground-based response functions. What is more, since the LIGO constraint on the speed of tensor modes $c_T$ is so stringent, in this section we will set $c_T = c$.

\subsection{Generalized ppE Scheme}

One of the generalization of the simplest ppE waveforms to theories with multiple polarizations can be written in the form~\cite{Chat12} \footnote{This is different from its original form of \cite{Chat12}, in order  to accommodate different propagation speeds, as mentioned above.},
\begin{widetext}
\bqn
\lb{FT3}
\tilde{h}(f) &=& \tilde{h}^{GR}(f)  \left(1+c_{ppE} \beta_{ppE} {\cal{U}}_2^{b_{ppE}+5}\right)e^{i 2 \beta_{ppE} {\cal{U}}_2^{b_{ppE}} } \nb\\
&& +\frac{{\cal{M}}^2}{R} {\cal{U}}_2^{-7/2} e^{i \Psi_{GR}^{(2)}} e^{i 2 \beta_{ppE} {\cal{U}}_2^{b_{ppE}} }  \left(1-\kappa_3^{1/2} c_{ppE} \beta_{ppE} {\cal{U}}_2^{b_{ppE}+5}\right)  \left[ \alpha_+ F_+ (1+\cos^2\vartheta)+\alpha_\times F_\times \cos \vartheta \right]\nb\\
&& +\frac{{\cal{M}}^2}{R} {\cal{U}}_2^{-7/2} e^{i \Psi_{GR}^{(2)}} e^{i 2 \beta_{ppE} {\cal{U}}_2^{b_{ppE}} } \left(1+ \kappa_3 c_{ppE} \beta_{ppE} {\cal{U}}_2^{b_{ppE}+5}\right)  \nb\\
&& ~~~\times \left \{ e^{i 2 \pi f R (1-c_S^{-1})} \left[\alpha_b F_b \sin^2 \vartheta+\alpha_L  F_L \sin^2 \vartheta \right] + e^{i 2 \pi f R (1-c_V^{-1})} \left[ \alpha_X F_X \sin (2\vartheta)+\alpha_Y  F_Y \sin \vartheta \right] \right \}\nb\\
&& +\eta^{1/5} \frac{{\cal{M}}^2}{R} {\cal{U}}_1^{-9/2} e^{i \Psi_{GR}^{(1)}} e^{i \beta_{ppE} {\cal{U}}_1^{b_{ppE}} }  \left(1+\kappa_3 c_{ppE} \beta_{ppE} {\cal{U}}_1^{b_{ppE}+5}\right)   \nb\\
&& ~~~\times \left \{ e^{i 2 \pi f R (1-c_S^{-1})} \left[ \gamma_b F_b \sin \vartheta+\gamma_L F_L \sin \vartheta \right] \right. \nb\\
&&~~~~~~~~~\left. + e^{i 2 \pi f R (1-c_V^{-1})} \left[\gamma_{X1} F_X \cos \vartheta+\gamma_{X2} F_X \sin \vartheta+\gamma_{Y1} F_Y+\gamma_{Y2} F_Y \sin \vartheta \right] \right \}, 
\eqn
where \cite{Chat12}
\bqn
\lb{hGRf}
\tilde{h}^{GR}(f) &=& -\sqrt{\frac{5 \pi}{96}} G_N^2  [F_+ (1+\cos^2 \vartheta)+2 i F_\times \cos \vartheta]  \frac{{\cal{M}}^2}{R} {\cal{U}}_2^{-7/2} e^{i \Psi^{(2)}_{GR}},
\eqn
\end{widetext}
and
\begin{align}
\lb{PsiGR}
 \Psi_{GR}^{(2)} &=  \frac{3}{128} (G_N \pi{\cal{M}}  f)^{-5/3}+ 2 \pi f \bar{t}_c  -2 \Phi(t_c)-\frac{\pi}{4}, \nb\\
 \Psi_{GR}^{(1)} &=  {\frac{3}{256} (2 G_N \pi {\cal{M}}  f)^{-5/3}}+2 \pi f \bar{t}_c   -\Phi(t_c)-\frac{\pi}{4},
\end{align}
with ${\cal{U}}_l \equiv (2 \pi G_N {\cal{M}} f/l)^{1/3}$. Note that $\varphi$ in \eqref{hGRf} has been set to zero to agree with those in \cite{Chat12,HYY15}.
 
Comparing Eq.~\eqref{eq:h-FT-SPA} with Eq.~\eqref{FT3} we see immediately that there is a mismatch. This is because the gravitational constants in $\ae$-theory that control binary motion are ${\cal{G}}$ and $G_{\ae}$, and thus, these constants appear in Eq.~\eqref{eq:h-FT-SPA}, while the ppE formalism is parameterized in terms of the gravitational constant observed on Earth $G_{N}$, which is why this constant appears in $\mathcal{U}_\ell$ in Eq.~\eqref{FT3}. The relation between ${\cal{G}}$ and $G_{N}$ is given explicitly in Eq.~\eqref{Gsen}, where we see that ${\cal{G}} = G_{N} + {\cal{O}}(s_{1},s_{2})$. Similarly, from Eq.~(\ref{Gae}) we see that  ${G_{\ae}} = G_{N} + {\cal{O}}(c_{14})$. The ppE formalism, however, is {\it{defined}} only in the limit of small deformations away from GR, and since $s_{1,2} \to 0$ and $c_{14} \to 0$ in the GR limit, one should really insert Eqs.~\eqref{Gsen} and~\eqref{Gae} into Eq.~\eqref{eq:h-FT-SPA}, then re-expand in small deformations, and then compare to Eq.~\eqref{FT3}, keeping only terms of leading-order in the coupling parameters and to leading-order in the PN approximation. To be specific, in the procedure of finding ppE parameters, we are going to apply the following approximations so that we can match Eqs.~\eqref{eq:h-FT-SPA} and \eqref{FT3}:
\bqn
\lb{Gappro}
& (1-c_{14})^{n_1} \left[(1-s_1)(1-s_2)\right]^{n_2} 
\nonumber \\
&=\left[1+{\cal{O}}(s_1, s_2)\right] \left[1+{\cal{O}}(c_{14})\right] \backsimeq 1,
\eqn
where $n_1$ and $n_2$ are arbitrary real numbers and the neglected contribution of ${\cal{O}}(s_1, s_2)$ and ${\cal{O}}(c_{14})$ enters at higher order in terms of the small coupling constants in the waveform.
 
The resulting Fourier transform of the response function in $\ae$-theory is still different from that in \cite{Chat12} because the former contains the factors of { $\exp[-2 \pi i f R(1-1/c_{N})]$ }discussed earlier. Therefore, in theories which contain additional polarization modes with different propagation speeds, we must generalize the results of  \cite{Chat12} by replacing every appearance of $F_{N}$ in Eq.~\eqref{FT3} {with $F_{N} \exp[-2 \pi i f R(1-1/c_{N})]$.} 

If we can re-cast Eq.~(\ref{FT1}) into the form of Eq.~(\ref{FT3}), then we can read off the set of ppE parameters \{$c_{ppE}$, $b_{ppE}$, $\beta_{ppE}$, $\alpha_+$, $\alpha_\times$, $\alpha_b$, $\alpha_L$, $\alpha_X$, $\alpha_Y$,
  $\gamma_b$, $\gamma_L$, $\gamma_{X1}$, $\gamma_{X2}$, $\gamma_{Y1}$, $\gamma_{Y2}$ \}.
 First, we observe that 
\bqn
\lb{ppEPsi12}
\Psi_{(2)} &=&   {\Psi^{(2)}_{GR}} +{\cal{U}}_2^{-7} \phi_1, \nb\\
\Psi_{(1)} &=&   {\Psi^{(1)}_{GR}} +\frac{1}{2}{\cal{U}}_1^{-7} \phi_1,
\eqn
where 
\bqn
\lb{phi12}
&&  {\phi_1} \equiv -\frac{3}{224} \eta^{2/5} \kappa_3^{-1} \epsilon_x,
\eqn
with
\bqn
\lb{kappa3}
{\kappa}_3 &\equiv& {\cal{A}}_1+{\cal{S}} {\cal{A}}_2+{\cal{S}}^2 {\cal{A}}_3.~~~ 
\eqn
Note that Eq.~\eqref{Gappro} has been used above and the $f$-dependent terms in $\phi_1$ are omitted to keep only the leading PN correction. With Eqs.~\eqref{Gappro} and~\eqref{ppEPsi12} at hand, we could write Eq.~(\ref{FT1}) as the desired form, i.e., Eq.~(\ref{FT3}). Here, we will omit the devilishly tedious expression for $\tilde{h}(f) $. Instead, we will first find the full expression of $\tilde{h}(f) $ and then read off the ppE parameters. The results are
 \allowdisplaybreaks
\bqn
\lb{ppE1}
&&c_{ppE} = \frac{224}{3}, \nb\\
&&b_{ppE} = -7, \nb\\
&&\beta_{ppE}  = \frac{1}{2} \phi_1 =  -\frac{3}{448} \kappa_3^{-1}\eta^{2/5} \epsilon_x, \nb\\
&&{\alpha_+} = { \frac{\sqrt{5 \pi}}{8 \sqrt{6}}  G_N^2 e^{i 2 \varphi}  \left(\kappa_3^{-1/2} -1  \right) g_+,} \nb\\
&&{\alpha_\times} ={-i  \frac{\sqrt{5 \pi}}{8 \sqrt{6}}   G_N^2  e^{i 2 \varphi}  \left( \kappa_3^{-1/2}-1 \right) g_\times,} \nb\\
&&{\alpha_b} = { \frac{\sqrt{5 \pi}}{8 \sqrt{6}} \kappa_3^{-1/2} G_N^2  e^{i 2 \varphi}  g_{b1},} \nb\\
&&{\alpha_L} = { \frac{\sqrt{5 \pi}}{8 \sqrt{6}} \kappa_3^{-1/2} G_N^2  e^{i 2 \varphi}  g_{L1},} \nb\\
&&{\alpha_X} = {  \frac{\sqrt{5 \pi}}{8 \sqrt{6}} \kappa_3^{-1/2} G_N^2  e^{i 2 \varphi}  g_{X1},} \nb\\
&&{\alpha_Y} = { -i  \frac{\sqrt{5 \pi}}{8 \sqrt{6}} \kappa_3^{-1/2} G_N^2 e^{i 2 \varphi}  g_{Y1},} \nb\\
&&{\gamma_b} = {-i  \frac{\sqrt{5 \pi}}{8 \sqrt{3}} \kappa_3^{-1/2} \eta^{-1/5} G_N^2  e^{i \varphi} (g_{b2}+g_{b4}),} \nb\\
&&{\gamma_L} = {-i \frac{\sqrt{5 \pi}}{8 \sqrt{3}} \kappa_3^{-1/2} \eta^{-1/5} G_N^2  e^{i \varphi} (g_{L2}+g_{L4}),} \nb\\
&&{\gamma_{X1}} = {-i \frac{\sqrt{5 \pi}}{8 \sqrt{3}} \kappa_3^{-1/2} \eta^{-1/5} G_N^2 e^{i \varphi} (g_{X2}+g_{X4}),} \nb\\
&&{\gamma_{X2}} = {-i \frac{\sqrt{5 \pi}}{8 \sqrt{3}} \kappa_3^{-1/2} \eta^{-1/5} G_N^2  e^{i \varphi} g_{X3},} \nb\\
&&{\gamma_{Y1}}= {\frac{\sqrt{5 \pi}}{8 \sqrt{3}} \kappa_3^{-1/2} \eta^{-1/5} G_N^2 e^{i \varphi} (g_{Y2}+g_{Y4}),} \nb\\
&&{\gamma_{Y2}} = {-i \frac{\sqrt{5 \pi}}{8 \sqrt{3}} \kappa_3^{-1/2} \eta^{-1/5} G_N^2 e^{i \varphi} g_{Y3},} 
\eqn
where \{$g_+$, $g_\times$, $g_{b1, 2, 4}$, $g_{L1, 2, 4}$, $g_{X1, 2, 3, 4}$, $g_{Y1, 2, 3, 4}$\} are functions given in Appendix D. Note that $g_{b2, 4}$, $g_{L2, 4}$, $g_{X2, 3, 4}$ and $g_{Y2, 3, 4} \propto \eta^{1/5}$. In other words, the $\eta$ terms in $\alpha_N$ and $\gamma_N$ actually have the same power, namely, $0$. Also note that $c_{ppE}$, which corresponds to the ratio between the amplitude and phase ppE corrections, agrees with that given in~\cite{Chat12,Tahura:2018zuq}.
 
 {Additionally, since the degenerate breathing and longitudinal modes,  in Eq.\eqref{FT3} we can put these terms together by introducing the 
quantities,}
\bqn
\lb{alphaS}
&&  { {\alpha_S} \equiv \alpha_b (1-a_{bL}),}
\eqn
  {where Eqs.\eqref{ggb}, \eqref{ppE1} and \eqref{RF2} had been used and $a_{bL}$ is given by \eqref{abL}.} 


\subsection{Fully Restricted ppE Approximation}

Now, we move to the regime of the fully restricted ppE approximation by mainly following \cite{HYY15}. We generalize~\cite{Chat12,HYY15} to allow for different propagation speeds of scalar, vector and tensor modes. This time, Eq.~(\ref{FT1}) is written in the form of

\bqn
\lb{FT5}
{\tilde{h}(f)}&=&   \sum_{N=S,V,T}\sum_{l=1}^{\infty}{A_{ppE}^{(l,N)}(f) e^{i \Psi_{ppE}^{(l,N)}(f)}}, 
\eqn
where
\bqn
\lb{ppEA}
{A_{ppE}^{(l,N)}(f) }&=& { A_{GR}^{(l)}(f) \left[ 1+ {\cal{U}}_l^{{\bar{a}}_{ppE}^{(l)}} \sum_{k=0}^{\infty} {\bar{\alpha}}_{ppE,k}^{(l,N)} ({\cal{U}}_l)^{k} \right],~~~~~~~~}\\
\lb{ppEPsi}
{\Psi_{ppE}^{(l,N)}(f)} &=&  \Psi_{GR}^{(l)}(f) + {\cal{U}}_l^{{\bar{b}}_{ppE}^{(l)}} \sum_{k=0}^{\infty} {\bar{\beta}}_{ppE,k}^{(l)} ({\cal{U}}_l)^{k} \nb \\ 
&&- 2\pi f R(1-c_N^{-1}).
\eqn
Here $l$ stands for the $l$-th harmonics and quantities with a GR subscript referring to expressions in the GR limits as in the last subsection.  {We also note that ${\bar{\alpha}}_{ppE,0}^{(l,N)} \neq 0$ and ${\bar{\beta}}_{ppE,0}^{(l)} \neq 0$, which means that the terms proportional to ${\cal{U}}_l^{{\bar{a}}_{ppE}^{(l)}}$ and ${\cal{U}}_l^{{\bar{b}}_{ppE}^{(l)}}$ correspond  to the term that enters at leading (lowest) PN order.} We can choose $c_T = 1$ since this effect has been absorbed by the redefinition of the coalescence time. Notice that when $c_S = c_V = c_T =1$, the phase is common to all of the scalar, vector and tensor modes, and the above formulation agrees with that in~\cite{Chat12}.
 
The restricted ppE waveform consists of amplitude corrections truncated to the leading PN order (which corresponds to $-1$ PN in our case) while phase corrections are kept to higher PN orders. In this paper, we consider the 
{\it fully restricted}  ppE waveform, in which  we only consider the dominant $l=2$ harmonic mode. Then, the above expressions can be reduced to

\bqn
\lb{FT6}
{\tilde{h}(f)}&\simeq& \sum_{N=S,V,T} { {A_{ppE}^{(2,N)}(f) e^{i \Psi_{ppE}^{(2,N)}(f)}}, }
\eqn
with
\bqn
\lb{ppEA2}
{A_{ppE}^{(2,N)}(f) }&=& { A_{GR}^{(2)}(f) \left[ 1+ {\cal{U}}_2^{{\bar{a}}_{ppE}^{(2)}}  {\bar{\alpha}}_{ppE, 0}^{(2, N)} \right],~~~~~~~~}\\
\lb{ppEPsi2}
{\Psi_{ppE}^{(2,N)}(f)} &=&  \Psi_{GR}^{(2)}(f) + {\cal{U}}_2^{{\bar{b}}_{ppE}^{(2)}} \sum_{k=0}^{\infty} {\bar{\beta}}_{ppE,k}^{(2)} ({\cal{U}}_2)^{k},  \nb \\ 
&&- 2\pi f R(1-c_N^{-1}).
\eqn
Here $\Psi_{GR}^{(2)}$ is given by Eq.~\eqref{PsiGR} while $A_{GR}^{(2)}$ is given by
\bqn
\lb{AGR2}
{A_{GR}^{(2)}} &=& - \sqrt{\frac{5\pi}{96}} \frac{\mathcal M^2}{R} (G_N \pi {\cal{M}} f)^{-7/6} G_N^2  \nb \\
&& \times \big[F_+ \big(1+\cos^2 \vartheta\big)+i 2 F_\times \cos \vartheta\big].
\eqn

Let us now determine the ppE parameters in Einstein-aether theory. 
Rewriting the waveform in Eq.~(\ref{eq:h-FT-SPA}) for the $l$=2 terms in a form given by Eq.~\eqref{FT6}, the ppE phase parameters can be extracted as
\bqn
\lb{ppE2}
&&{{\bar{b}}_{ppE}^{(2)}} = {-7,} \nb\\
&&{{\bar{\beta}}_{ppE, 0}^{(2)}}=  {\phi_1}= {-\frac{3}{224} \kappa_3^{-1} \eta^{2/5}  \epsilon_x, } \nb\\
&&{{\bar{\beta}}_{ppE, 1}^{(2)}} =  {0,} \nb\\
	&&  {{\bar{\beta}}_{ppE, 2}^{(2)}}= -{\frac{3}{128} \left[- \frac{2}{3} \left( s_1+ s_2\right)-\frac{1}{2} c_{14} +(\kappa_3-1) \right]}. \nb\\
\eqn
Notice that ${{\bar{\beta}}_{ppE, 0}^{(2)}}$ is different from $\beta_{ppE}^{(2)}$ in Eq.~\eqref{ppE1} by a factor of 2 due to a prefactor 2 in front of $\beta_{ppE}^{(2)}$ in Eq.~\eqref{FT3}. When deriving ${{\bar{\beta}}_{ppE, 2}^{(2)}}$, we kept ${\cal{O}}(s_1, s_2, c_{14})$ contribution in Eq.~\eqref{Gappro} for consistency. Next, the ppE amplitude parameters are extracted as
\bqn
\lb{ppE3}
{{\bar{a}}_{ppE}^{(2)}} &=& {-2,} \nb\\
{{\bar{\alpha}}_{ppE, 0}^{(2,T)}} &=& -\frac{1}{2} \kappa_3^{-1/2} \eta^{2/5} \epsilon_x, \nb \\
{{\bar{\alpha}}_{ppE, 0}^{(2,S)}} &=& {{\bar{\alpha}}_{ppE, 0}^{(2,T)}}\frac{g_{b1} F_b \sin^2 \vartheta + g_{L1} F_L \sin^2 \vartheta}{g_+ F_+\big (1+\cos^2 \vartheta \big)-i g_\times F_\times \cos \vartheta}\nb \\
&& {= {{\bar{\alpha}}_{ppE, 0}^{(2,T)}}\frac{g_{b1} F_b \sin^2 \vartheta (1-a_{bL})}{g_+ F_+\big (1+\cos^2 \vartheta \big)-i g_\times F_\times \cos \vartheta}, } \nb\\
{{\bar{\alpha}}_{ppE, 0}^{(2,V)}} &=& {{\bar{\alpha}}_{ppE, 0}^{(2,T)}}\frac{g_{X1} F_X \sin (2\vartheta)-i  g_{Y1} F_Y \sin \vartheta}{g_+ F_+\big (1+\cos^2 \vartheta \big)-i g_\times F_\times \cos \vartheta}. \nb \\
\eqn
The above $F_N$ and $\vartheta$ dependence on the ppE amplitude parameters for the scalar and vector modes seem to be a generic feature, as predicted in~\cite{Chat12}. We note that even if the denominator $g_+ F_+\big (1+\cos^2 \vartheta \big)-i g_\times F_\times \cos \vartheta$ in ${{\bar{\alpha}}_{ppE, 0}^{(2,S)}}$ and ${{\bar{\alpha}}_{ppE, 0}^{(2,V)}}$ becomes 0, the scalar and vector mode corrections to the waveform amplitude do not diverge since the ppE parameters are multiplied by ${A_{GR}^{(2)}}$, which contains the same factor that cancels the denominator of ${{\bar{\alpha}}_{ppE, 0}^{(2,S)}}$ and ${{\bar{\alpha}}_{ppE, 0}^{(2,V)}}$.

Let us now compare the results presented here against those in~\cite{HYY15,Tahura:2018zuq}. 
First, ${{\bar{b}}_{ppE}^{(2)}}$ agrees with that in~\cite{HYY15,Tahura:2018zuq}, while ${{\bar{a}}_{ppE}^{(2)}}$ agrees with that in \cite{Tahura:2018zuq}, which corrected~\cite{HYY15}. Second, in~\cite{HYY15}, the aether field is assumed to be aligned with the CMB frame and $V \sim 10^{-3}$, which is much slower than the relative velocity of the binary constituents before coalescence. In this case, the dominant contribution in $\epsilon_x$ in Eq.~\eqref{epsilonx} arises from the term proportional to $\mathcal C$. Moreover, the denominator ${\cal{A}}_1+{\cal{S}} {\cal{A}}_2+{\cal{S}}^2 {\cal{A}}_3$ originates from factoring out the 0PN contribution in $\dot \omega_s$ in Eq.~\eqref{omegas1}. If we neglect the Einstein-aether correction at 0PN order, this factor can be simply set to the GR value of 1 (and one can take the similar limit in $\kappa_3$). Then, the leading ppE phase ${{\bar{\beta}}_{ppE, 0}^{(2)}}$ in Eq.~\eqref{ppE2} agrees with that in~\cite{HYY15,Tahura:2018zuq} within the approximation in Eq.~\eqref{Gappro}.
Similarly, ${{\bar{\alpha}}_{ppE, 0}^{(2)}}$ reduces to the leading ppE amplitude correction in~\cite{Tahura:2018zuq} under the small coupling approximation. On the other hand, ${{\bar{\beta}}_{ppE, 2}^{(2)}}$ in Eq.~\eqref{ppE2} corrects that in~\cite{HYY15}.

\section{Conclusions}
 \renewcommand{\theequation}{6.\arabic{equation}} \setcounter{equation}{0}

In this paper, we  have studied the  waveforms and polarizations of GWs emitted by a binary system in Einstein-aether theory, which contains four dimensionless coupling parameters $c_i$'s. We focused on the inspiral phase, adopted the PN approximations and assumed that the Einstein-aether coupling constants are small. In $\ae$-theory, 
all the six polarization modes of GWs, referred to as $h_N (N = +, \times, b, L,  {X, Y})$, are present, although only five of them are independent, as  the breathing and longitudinal  modes ($h_b$ and $h_L$) are  proportional 
to each other. In the GR limit of $c_i \to 0 \; (i = 1, 2, 3, 4)$, only the ``+" and ``$\times$"  modes remain and they reduce to those  of GR as expected. 

Gravitational waveforms and GW polarizations emitted by a binary system in the inspiral phase in  $\ae$-theory were already studied in \cite{HYY15}. In the current paper, we have first re-derived these
formulas, and corrected some typos, by keeping all the terms to ${\cal{O}}(v^2)$. In particular, we have shown explicitly that the non-relativistic GW modes $h_{b, L,  {X, Y}}$ contain not 
only the first harmonic terms of the orbital phase, as shown in 
\cite{HYY15}, but also the second harmonic ones when one includes higher PN order terms. 
  
 Note also that in deriving the expressions of the GW polarization modes $h_N$'s [{cf.} (\ref{hp})-(\ref{hy})], we have not assumed that COM of the binary system is always comoving with the aether field. 
 In fact, in cosmology the aether field is normally assumed to be comoving with CMB \cite{Jacobson}. As a result, individual compact objects in the Universe, such as galaxies and massive stars,
 are in general expected to have peculiar velocities with respect to the CMB. A typical velocity of compact objects in our own galaxy in this frame is about $V^2 \simeq 10^{-6}$, for which Foster had shown that the PN
 approximations adopted here are valid \cite{Foster07}.

 Using the SPA method \cite{Chat12,YBW09,Tan18}, we have also calculated the response function and its Fourier transform  for both ground- and space-based GW detectors. 
We then generalized the ppE framework to allow for different propagation speeds among scalar, vector and tensor modes. The ppE parameters within this new framework is given
 by Eqs.~{(\ref{ppE1}), (\ref{ppE2}) and (\ref{ppE3})}, which depend on all six polarization modes. The leading ppE phase correction at $-1$PN order agrees with that in \cite{HYY15,Tahura:2018zuq} under the small coupling approximation and within the CMB frame. Similarly, the leading ppE amplitude correction agrees with that in~\cite{Tahura:2018zuq} under the same approximation. On the other hand, the next-to-leading ppE correction in the phase at 0PN order corrects the corresponding expression in~\cite{HYY15}.

\section*{Acknowledgments}

A.Z., C.Z. and X.Z. would like to express their gratitude to ITPC for hospitality.  
N.~Y~ would like to thank Enrico Barausse and Thomas Sotiriou for several conversations that clarified the effect of the position of the center of mass in Einstein-aether calculations. 
We would also like to thank K. Lin, T. Liu, R. Niu, S.-J. Zhang, and X. Zhang for their valuable discussions. 
A.Z., C.Z. and X.Z acknowledge support in part by the National Natural Science
Foundation of China (NNSFC) with the grant numbers:  Nos. 11603020, 11633001, 11173021, 11322324, 11653002, 11421303, 11375153, 11675145, 
11675143,  11105120, 11805166, 11835009, 11690022, 11375247, 11435006, 11575109,  11647601, and No. 11773028.
K.Y. acknowledges support from NSF Award PHY-1806776, a Sloan Foundation Research Fellowship, the Ed Owens Fund, the COST Action GWverse CA16104 and JSPS KAKENHI Grants No. JP17H06358.
N.Y.   acknowledges support from NSF grant PHY-1759615, and NASA grants 80NSSC18K1352.


\section*{Appendix A: The stationary phase approximation}
\renewcommand{\theequation}{A.\arabic{equation}}
\setcounter{equation}{0}

SPA is a useful method  for dealing with the Fourier transform  (FT) of the response functions. The details of this method can be found in \cite{Chat12, YBW09, Tan18}. 
Here we will provide a brief introduction to this technique.

For real $g_0(t)$, $\psi_0(t)$, $a_0$, $b_0$ and $y_0$, we have the following approximation to $g_0(t)$'s Fourier integral \cite{Olver, Orszag},
\bqn
\lb{approx}
\lim_{y_0\to\infty} I_0(y_0) &\equiv& \lim_{y_0\to\infty} \int_{a_0}^{b_0} g_0(t) e^{i y_0 \psi_0(t)} dt \nb\\
&\approx& \lim_{y_0\to\infty} g_0(t_a) e^{i y_0 \psi_0(t_a) \pm \frac{i \pi}{2 l}} \nb\\
&&\times \left[\frac{l!}{y_0 |\psi_0^{(l)}(t_a)|} \right]^{1/l} \frac{\Gamma(1/l)}{l},
\eqn
where $\psi_0^{(l)}(t)$ denotes the $l$-th derivative with respect to $t$.  $\Gamma(x)$ denotes the   Gamma function   \cite{Bayin}.
$t_a$ refers to the stationary point that is determined by the conditions
\bqn
\lb{ta}
&& \psi_0^{(1)}(t_a) = \psi_0^{(2)}(t_a) = ... = \psi_0^{(l-1)}(t_a) = 0, \nb\\
&& \psi_0^{(l)}(t_a) \ne 0,
\eqn
and we will choose ``+" for (\ref{approx}) when $\psi_0^{(l)}(t_a) > 0$,  and  ``-" for (\ref{approx}) when $\psi_0^{(l)}(t_a) < 0$. Besides, the validity of this approximation requires
\bqn
\lb{g0t}
\left |\int_{a_0}^{b_0} g_0(t) dt \right| < \infty,
\eqn
and $\psi_0(t)$ is not a constant on any interval $U_0 \in [a_0, b_0]$. 
As an example, we will use  SPA to calculate the FT for the response function,
\bqn
\lb{Hnt}
H_n (t) &=& q_n \omega_s^{2/3}(t_r) \cos (2 \Phi(t_r)),
\eqn
where $t_r=t- R/v_s$ is the retarded time with $v_s$ denoting the speed of the wave. 

To make sure that the approximation (\ref{approx}) is valid for the calculation of the FT of (\ref{Hnt}), we need to assume that $d[\ln(q_n \omega_s^{2/3})]/dt \ll d \Phi/dt$ and $d^2 \Phi/dt^2 \ll (d \Phi/ dt)^2$. 
 Then, using (\ref{FT1}) and Euler's formula, we find
\bqn
\lb{Hnf}
\tilde{H}_n (f) &=& \frac{1}{2}q_n e^{i 2 \pi f  R/v_s}  \nb\\
&&\times \int \omega_s^{2/3} \left[e^{i (-2 \Phi+2 \pi f t)} +e^{i (2 \Phi+2 \pi f t)} \right] dt. \nb\\
\eqn
Since $d (2 \Phi+2 \pi f t)/dt=0$, we find  $ \dot{\Phi}(t_a) = - \pi f$,  which leads to a non-physical frequency $f$ and thus can be discarded. Conversely, from the first term in (\ref{Hnf}), we find  
$\dot{\Phi}(t_a)=\pi f$ by $d (-2 \Phi+2 \pi f t)/dt |_{t_a}=0$.  Thus, we obtain $\omega_s(t_a)=\dot{\Phi}(t_a)=\pi f$ and $l=2$ for (\ref{approx}). Now we write $\tilde{H}_n (f)$ as
\bqn
\lb{Hnf2}
\tilde{H}_n (f) &=& { \frac{1}{2}q_n e^{i 2 \pi f  R/v_s}   } I_n(f),
\eqn
where
\bqn
\lb{In}
I_n (f) &\equiv& \int \omega_s^{2/3} \left[e^{i (-2 \Phi+2 \pi f t)} \right] dt.
\eqn
Note that there is no summation in (\ref{Hnf2}) with respect to $n$. At the same time, from (\ref{omegas1}) we find that $d^2 (-2 \Phi+2 \pi f t)/dt^2 |_{t_a} = -2 \ddot{\Phi}(t_a) = -2 \dot{\omega}(t_a) 
\sim -{\omega}^{11/3}(t_a) < 0$,  which helps us to determine the sign in (\ref{approx}). With all of these in hand, we can apply the approximation (\ref{approx}) to (\ref{In}),  and find that
\bqn
\lb{In2}
I_n (f) &\simeq& \frac{1}{2} \omega_s^{2/3}(t_a) \sqrt{\frac{\pi}{\dot{\omega}_s(t_a)}} \times 2 e^{i f \psi_n- \frac{i \pi}{4}},
\eqn
where
\bqn
\lb{psin}
\psi_n (t) &\equiv& -\frac{2 \Phi(t)}{f}+2 \pi t.
\eqn
Note that in the above expression there is an additional factor of $2$, which originates from the analysis of \cite{YBW09}. Substituting (\ref{In2}) into (\ref{Hnf2}), we find
\bqn
\lb{Hnf3}
\tilde{H}_n (f) &=& \frac{\sqrt{\pi}}{2} q_n \left[ \omega_s^{2/3}(t_a) \dot{\omega}_s^{-1/2}(t_a) \right] e^{i \Psi_n}, 
\eqn
where
\bqn
\lb{Psin}
\Psi_n &\equiv& -2 \Phi(t_a)+2 \pi f t_a+2 \pi f \frac{R}{v_s} -\frac{\pi}{4}.
\eqn

Next, using the relation
\bqn
\lb{int1}
[-2 \Phi(t)+2 \pi f t]|_{t_c}^{t_a} &=& \int_{t_c}^{t_a} \frac{d[-2 \Phi(t)+2 \pi f t]}{dt}dt, \nb\\
\eqn
and the fact that $\omega_s(t_c) \to \infty$, we can carry out the integral on the right-hand side of (\ref{int1}) approximately,  and finally obtain 
\bqn
\lb{Psin2}
\Psi_n &=& \frac{9}{20} \kappa_1^{-1} ({\cal{G}} \pi m  f)^{-5/3} \left[1-\frac{4}{7}({\cal{G}} \pi m  f)^{-2/3}\epsilon_x \right] \nb\\
&&+2 \pi f \left( t_c+\frac{R}{v_s} \right)-2 \Phi(t_c)-\frac{\pi}{4},
\eqn
where the asymptotical form of the $\dot{\omega}_s$ and  $\ddot{\omega}_s$ had been used.

Similarly,  using the relation
\bqn
\lb{int2}
[\omega_s^{2/3}(t) \dot{\omega}_s^{-1/2}(t) ]|_{t_c}^{t_a} &=& \int_{t_c}^{t_a} \frac{d[ \omega_s^{2/3}(t) \dot{\omega}_s^{-1/2}(t) ]}{dt}dt, \nb\\
\eqn
and $\omega_s(t_c) \to \infty$, we can also carry out the integral on the right-hand side of (\ref{int2}).
Finally, we find
\bqn
\lb{Hnf4}
\tilde{H}_n (f) &=& \frac{\sqrt{\pi}}{2} ({\cal{G}} m)^{1/3} q_n \kappa_1^{-1/2} ({\cal{G}} \pi mf)^{-7/6} \nb\\
&& \times \left[1-\frac{1}{2} ({\cal{G}} \pi m  f)^{-2/3} \epsilon_x  \right] e^{i \Psi_n}, ~~~~
\eqn
where $\Psi_n$ is given by (\ref{Psin2}). The  calculations for (\ref{RF5}) can be obtained by following the same steps\footnote{Of course, there is a difference between the demonstration here and the calculations in Sec. V. That is, in Sec. V, the phase \eqref{Psin2} is fixed for the 1st and 2nd harmonic terms. At the same time, the term that related to $v_s$ is absorbed into the amplitude part. Logically, this seems to be a big change. Nevertheless, mathematically, this modification is actually trivial.}.

\section*{Appendix B: The Expressions of $q_{N (l)}$}
\renewcommand{\theequation}{B.\arabic{equation}}
\setcounter{equation}{0}

In (\ref{RF5}) we introduced $q_{N (l)}$, which are given explicitly by
\bqn
\lb{qq}
&& q_{+ (1)} \equiv d_{+} \cos(2 \varphi) F_+, \nb\\
&& q_{+ (2)} \equiv d_{+} \sin(2 \varphi) F_+, \nb\\
&& q_{+ (3)} = q_{+ (4)}   = 0, \nb\\
&& q_{\times (1)} \equiv d_{\times} \sin(2 \varphi) F_\times, \nb\\
&& q_{\times (2)} \equiv -d_{\times} \cos(2 \varphi) F_\times, \nb\\
&& q_{\times (3)} = q_{\times (4)}  = 0, \nb\\
&& q_{b (1)} \equiv d_{b1} \cos(2 \varphi) F_b, \nb\\
&&  q_{b (2)} \equiv d_{b1} \sin(2 \varphi) F_b, \nb\\
&& q_{b (3)} \equiv (d_{b2}+d_{b4}) \sin \varphi F_b, \nb\\
&& q_{b (4)} \equiv -(d_{b2}+d_{b4}) \cos \varphi F_b, \nb\\
&& q_{L (1)} \equiv d_{L1} \cos(2 \varphi) F_L, \nb\\
&& q_{L (2)} \equiv d_{L1} \sin(2 \varphi) F_L, \nb\\
&& q_{L (3)} \equiv (d_{L2}+d_{L4}) \sin \varphi F_L, \nb\\
&& q_{L (4)} \equiv -(d_{L2}+d_{L4}) \cos \varphi F_L, \nb\\
&& q_{{ {X}} (1)} \equiv d_{{ {X}}1} \cos(2 \varphi) F_{ {X}}, \nb\\
&&  q_{{ {X}} (2)} \equiv d_{{ {X}}1} \sin(2 \varphi) F_{ {X}}, \nb\\
&& q_{{ {X}} (3)} \equiv (d_{{ {X}}2}+d_{{ {X}}4}) \sin \varphi F_{ {X}}, \nb\\
&& q_{{ {X}} (4)} \equiv -(d_{{ {X}}2}+d_{{ {X}}4}) \cos \varphi F_{ {X}}, \nb\\
&& q_{{ {Y}} (1)} \equiv d_{{ {Y}}1} \sin(2 \varphi) F_{ {Y}}, \nb\\
&& q_{{ {Y}} (2)} \equiv -d_{{ {Y}}1} \cos(2 \varphi) F_{ {Y}}, \nb\\
&& q_{{ {Y}} (3)} \equiv [(d_{{ {Y}}2}+d_{{ {Y}}4}) \cos \varphi+ d_{{ {Y}}5} \sin \varphi] F_{ {Y}}, \nb\\
&& q_{{ {Y}} (4)} \equiv [(d_{{ {Y}}2}+d_{{ {Y}}4}) \sin \varphi- d_{{ {Y}}5} \cos \varphi] F_{ {Y}},  
\eqn
 where
\bqn
\lb{dd}
d_+ &\equiv& -\frac{2 G_{\ae}}{ R} {\cal{G}}^{2/3} {\cal{M}}^{5/3} (1+\cos^2 \vartheta), \nb\\
d_\times &\equiv& \frac{4 G_{\ae}}{ R} {\cal{G}}^{2/3} {\cal{M}}^{5/3} \cos \vartheta, 
\eqn
\bqn
\lb{ddb}
d_{b1} &\equiv& \frac{2 G_{\ae}}{ R} \frac{c_{14}}{2-c_{14}} \frac{-3 c_{14} (Z-1) c_S^2+2 \cal{S}}{c_{14} c_S^2} \nb\\
&& \times {\cal{G}}^{2/3} {\cal{M}}^{5/3} \sin^2 \vartheta, \nb\\
d_{b2} &\equiv& \frac{2 G_{\ae}}{ R} \frac{c_{14}}{2-c_{14}} \frac{2 \Delta s}{c_{14} c_S} \eta^{1/5} {\cal{G}}^{1/3} {\cal{M}}^{4/3} \sin \vartheta, \nb\\
d_{b4} &\equiv& -\frac{2 G_{\ae}}{ R} \frac{c_{14}}{2-c_{14}} \frac{4 \Delta s}{c_{14} c_S^2} \eta^{1/5} {\cal{G}}^{1/3} {\cal{M}}^{4/3} \sin \vartheta N^i V^i, \nb\\
d_{L1} &\equiv& a_{bL} d_{b1}, \nb\\
d_{L2} &\equiv& a_{bL} d_{b2}, \nb\\
d_{L4} &\equiv& a_{bL} d_{b4}, 
\eqn
\bqn
\lb{ddc}
d_{{ {X}}1} &\equiv& -\frac{\beta_1 G_{\ae}}{ R} \frac{1}{2 c_1-c_{13}  c_-} \frac{1} {c_V} \nb\\
&& \times  \left( {\cal{S}}-\frac{c_{13} }{1-c_{13} } \right) {\cal{G}}^{2/3} {\cal{M}}^{5/3} \sin (2 \vartheta), \nb\\
d_{{ {X}}2} &\equiv& 2 \frac{\beta_1 G_{\ae}}{ R} \frac{1}{2 c_1-c_{13}  c_-} \Delta s \eta^{1/5} {\cal{G}}^{1/3} {\cal{M}}^{4/3} \cos \vartheta, \nb\\
d_{{ {X}}4} &\equiv& \frac{\beta_1 G_{\ae}}{ R} \frac{1}{2 c_1-c_{13}  c_-} \frac{2 \Delta s}{c_V} \eta^{1/5} {\cal{G}}^{1/3} {\cal{M}}^{4/3} \nb\\
&& \times (\sin \vartheta e_X^i+\cos \vartheta N^i) V^i, \nb\\
d_{{ {Y}}1} &\equiv& \frac{\beta_1 G_{\ae}}{ R} \frac{1}{2 c_1-c_{13}  c_-} \frac{2} {c_V} \nb\\
&& \times  \left( {\cal{S}}-\frac{c_{13} }{1-c_{13} } \right) {\cal{G}}^{2/3} {\cal{M}}^{5/3} \sin \vartheta, \nb\\
d_{{ {Y}}2} &\equiv& 2 \frac{\beta_1 G_{\ae}}{ R} \frac{1}{2 c_1-c_{13}  c_-} \Delta s \eta^{1/5} {\cal{G}}^{1/3} {\cal{M}}^{4/3}, \nb\\
d_{{ {Y}}4} &\equiv& \frac{\beta_1 G_{\ae}}{ R} \frac{1}{2 c_1-c_{13}  c_-} \frac{2 \Delta s}{c_V} \eta^{1/5} {\cal{G}}^{1/3} {\cal{M}}^{4/3} N^i V^i, \nb\\
d_{{ {Y}}5} &\equiv& \frac{\beta_1 G_{\ae}}{ R} \frac{1}{2 c_1-c_{13}  c_-} \frac{2 \Delta s}{c_V} \eta^{1/5} {\cal{G}}^{1/3} {\cal{M}}^{4/3} \nb\\
&& \times \sin \vartheta e_Y^i V^i, 
\eqn
and
\bqn
\lb{abL}
a_{bL} &\equiv& 1+2 \beta_2.
\eqn
Note that the all  $d_{{ {X}}}$'s and $d_{{ {Y}}}$'s are proportional to $\beta_1$, and therefore proportional to $c_{13}$.

\section*{Appendix C: The Expressions of $q^{\prime}_{N (l)}$}
\renewcommand{\theequation}{C.\arabic{equation}}
\setcounter{equation}{0}

In (\ref{RFL2}) we introduced $q^{\prime}_{N (l)}$'s,  which are given by
\bqn
\lb{qqL}
&& q^{\prime}_{+ (1)} \equiv d_{+} \cos(2 \varphi) F^{\prime}_+(t), \nb\\
&& q^{\prime}_{+ (2)} \equiv d_{+} \sin(2 \varphi) F^{\prime}_+(t), \nb\\
&& q^{\prime}_{+ (3)} = q^{\prime}_{+ (4)}   = 0, \nb\\
&& q^{\prime}_{\times (1)} \equiv d_{\times} \sin(2 \varphi) F^{\prime}_\times (t),  \nb\\
&& q^{\prime}_{\times (2)} \equiv -d_{\times} \cos(2 \varphi) F^{\prime}_\times (t), \nb\\
&& q^{\prime}_{\times (3)} = q^{\prime}_{\times (4)}  = 0, \nb\\
&& q^{\prime}_{b (1)} \equiv d_{b1} \cos(2 \varphi) F^{\prime}_b (t), \nb\\
&&  q^{\prime}_{b (2)} \equiv d_{b1} \sin(2 \varphi) F^{\prime}_b (t), \nb\\
&& q^{\prime}_{b (3)} \equiv (d_{b2}+d_{b4}) \sin \varphi F^{\prime}_b (t), \nb\\
&& q^{\prime}_{b (4)} \equiv -(d_{b2}+d_{b4}) \cos \varphi F^{\prime}_b (t), \nb\\
&& q^{\prime}_{L (1)} \equiv d_{L1} \cos(2 \varphi) F^{\prime}_L (t), \nb\\
&& q^{\prime}_{L (2)} \equiv d_{L1} \sin(2 \varphi) F^{\prime}_L (t), \nb\\
&& q^{\prime}_{L (3)} \equiv (d_{L2}+d_{L4}) \sin \varphi F^{\prime}_L (t), \nb\\
&& q^{\prime}_{L (4)} \equiv -(d_{L2}+d_{L4}) \cos \varphi F^{\prime}_L (t), \nb\\
&& q^{\prime}_{{ {X}} (1)} \equiv d_{{ {X}}1} \cos(2 \varphi) F^{\prime}_{ {X}} (t), \nb\\
&&  q^{\prime}_{{ {X}} (2)} \equiv d_{{ {X}}1} \sin(2 \varphi) F^{\prime}_{ {X}} (t), \nb\\
&& q^{\prime}_{{ {X}} (3)} \equiv (d_{{ {X}}2}+d_{{ {X}}4}) \sin \varphi F^{\prime}_{ {X}} (t), \nb\\
&& q^{\prime}_{{ {X}} (4)} \equiv -(d_{{ {X}}2}+d_{{ {X}}4}) \cos \varphi F^{\prime}_{ {X}} (t), \nb\\
&& q^{\prime}_{{ {Y}} (1)} \equiv d_{{ {Y}}1} \sin(2 \varphi) F^{\prime}_{ {Y}} (t), \nb\\
&& q^{\prime}_{{ {Y}} (2)} \equiv -d_{{ {Y}}1} \cos(2 \varphi) F^{\prime}_{ {Y}} (t), \nb\\
&& q^{\prime}_{{ {Y}} (3)} \equiv [(d_{{ {Y}}2}+d_{{ {Y}}4}) \cos \varphi+ d_{{ {Y}}5} \sin \varphi] F^{\prime}_{ {Y}} (t), \nb\\
&& q^{\prime}_{{ {Y}} (4)} \equiv [(d_{{ {Y}}2}+d_{{ {Y}}4}) \sin \varphi- d_{{ {Y}}5} \cos \varphi] F^{\prime}_{ {Y}} (t),   
\eqn
where $d_{Nl}$ are given by {(\ref{dd}-\ref{ddc})} and, 
\bqn
\lb{Ft}
F^{\prime}_+ (t) &\equiv& \frac{1}{2} [1+\cos^2 \theta(t) ] \sin [2 \phi(t)] \cos [2 \psi(t)] \nb\\
&&+\cos [\theta(t)] \cos [2 \phi(t)] \sin [2 \psi(t)], \nb\\
F^{\prime}_\times (t) &\equiv& \frac{1}{2} {[}1+\cos^2 \theta(t)] \sin [2 \phi(t)] \sin [2 \psi(t)] \nb\\
&&-\cos [\theta(t)] \cos [2 \phi(t)] \cos [2 \psi(t)], \nb\\
F^{\prime}_b (t) &\equiv& -\frac{1}{2} \sin^2 [\theta(t)] \sin [2 \phi(t)], \nb\\
F^{\prime}_L (t) &\equiv& \frac{1}{2} \sin^2 [\theta(t)] \sin [2 \phi(t)], \nb\\
F^{\prime}_{ {X}} (t) &\equiv& - \sin [\theta(t)] 
\{\cos [\theta(t)] \sin [2 \phi(t)] \cos [\psi(t)] \nb\\
&&+\cos [2 \phi(t)] \sin [\psi(t) \}, \nb\\
F^{\prime}_{ {Y}} (t) &\equiv& \sin [\theta(t)]  
{\{} {-\cos [\theta(t)] \sin [2 \phi(t)]} \sin [\psi(t)] \nb\\
&& +\cos [2 \phi(t)] \cos [\psi(t)] \}.
\eqn

The angles  $\theta(t)$, $\phi(t)$ and $\psi(t)$   are given by
\bqn
\lb{tpp}
\theta(t) &=& \cos^{-1}\left \{\frac{1}{2} [\cos{\bar \theta}-\sqrt{3} \cos(\bar \phi-\bar \Phi) \sin \bar \theta] \right\}, \nb\\
\phi(t) &=& -\tan^{-1}\Bigg \{\frac{1}{2} \csc \bar \theta \csc (\bar \phi-\bar \Phi) [\sqrt{3} \cos \bar \theta \nb\\
&&  +\cos (\bar \phi-\bar \Phi) \sin \bar \theta] \Bigg\}+\Lambda, \nb\\
\psi(t) &=& -\tan^{-1}\Bigg \{[\sqrt{3} \cos \bar \phi \nb\\
&&  {\times(\cos \bar \psi \sin \bar \Phi-\cos \theta \sin \bar \psi \cos \bar \Phi)} \nb\\
&& -\sin \bar \psi (\sin \bar \theta+\sqrt{3} \cos \bar \theta \sin \bar \phi \sin \bar \Phi) \nb\\
&& -\sqrt{3} \sin \bar \phi \cos \bar \psi \cos \bar \Phi]  \nb\\
&&\times  [\sqrt{3} (\cos \bar \theta \cos \bar \phi \cos \bar \psi-\sin \bar \phi \sin \bar \psi) \cos \bar \Phi \nb\\
&& +\cos \bar \psi (\sin \bar \theta+\sqrt{3} \cos \bar \theta \sin \bar \phi \sin \bar \Phi) \nb\\
&& +\sqrt{3} \cos \bar \phi \sin \bar \psi \sin \bar \Phi]^{-1} \Bigg\}, 
\eqn
where \cite{Cut98}
\bqn
\lb{Lambda}
\Lambda &=& \Lambda_0+\frac{2 \pi t}{T_0},
\eqn
which is the phase for the rotation of the three satellites around the COM of LISA with $\Lambda_0$ being a constant,
 and $\bar \Phi$ is provided in (\ref{Phibar}). Just like in (\ref{Phibar}), $T_0$ here is equal to the sidereal period of the Earth. Here $\bar \theta$, $\bar \phi$ and $\bar \psi$ are the  {three angles related to the center of the binary with respect to the Sun (note that their definitions are different from the general Euler angles \cite{Gold}), defined explicitly  in \cite{Cut98} and Sec.~11.5 of \cite{PW14}}, and can be treated as constants. 
 Note that once the detector is specified, e.g. LISA, $\varphi$ and $\vartheta$ in $q^{\prime}_{N(l)}$ will be determined by $\{\bar \theta, \bar \phi, \bar \psi\}$, i.e.  
 $\{\bar \theta, \bar \phi, \bar \psi, \vartheta, \varphi\}$ are not independent.

\section*{Appendix D: The Expressions of $g_{N}$}
\renewcommand{\theequation}{D.\arabic{equation}}
\setcounter{equation}{0}

{In (\ref{ppE1}) we use the factors $g_N$ ($g_N$ $\in$ \{$g_+$, $g_\times$, $g_{b1, 2, 4}$, $g_{L1, 2, 4}$, $g_{X1, 2, 3, 4}$, $g_{Y1, 2, 3, 4}$ \}), which are given as follows:}

\bqn
\lb{gga}
{g_+ \equiv -2, \quad  \quad   \quad   \quad   g_\times \equiv 4,}
\eqn
\bqn
\lb{ggb}
{g_{b1} } &\equiv& {\frac{2 c_{14}}{2-c_{14}} \frac{-3 c_{14} (Z-1) c_S^2+2 \cal{S}}{c_{14} c_S^2}, }\nb\\
{g_{b2} } &\equiv& {\frac{2 c_{14}}{2-c_{14}} \frac{2 \Delta s}{c_{14} c_S} \eta^{1/5} ,} \nb\\
{g_{b4} } &\equiv& {- \frac{2 c_{14}}{2-c_{14}} \frac{4 \Delta s}{c_{14} c_S^2} \eta^{1/5}  N^i V^i,} \nb\\
{g_{L1}} &\equiv& {a_{bL} g_{b1},} \nb\\
{g_{L2}} &\equiv& {a_{bL} g_{b2},} \nb\\
{g_{L4}} &\equiv& {a_{bL} g_{b4},} 
\eqn
\bqn
\lb{ggc}
{g_{{ {X}}1}} &\equiv& {-\frac{\beta_1}{2 c_1-c_{13}  c_-} \frac{1} {c_V} \left( {\cal{S}}-\frac{c_{13} }{1-c_{13} } \right),} \nb\\
{g_{{ {X}}2}} &\equiv& { \frac{2 \beta_1}{2 c_1-c_{13}  c_-} \Delta s \eta^{1/5},} \nb\\
{g_{{ {X}}3}} &\equiv& {\frac{\beta_1}{2 c_1-c_{13}  c_-} \frac{2 \Delta s}{c_V} \eta^{1/5} e_X^i V^i,} \nb\\
{g_{{ {X}}4}} &\equiv& {\frac{\beta_1}{2 c_1-c_{13}  c_-} \frac{2 \Delta s}{c_V} \eta^{1/5}  N^i V^i,} \nb\\
{g_{{ {Y}}1}} &\equiv& {\frac{\beta_1}{2 c_1-c_{13}  c_-} \frac{2} {c_V} \left( {\cal{S}}-\frac{c_{13} }{1-c_{13} } \right),} \nb\\
{g_{{ {Y}}2}} &\equiv& { \frac{2 \beta_1}{2 c_1-c_{13}  c_-} \Delta s \eta^{1/5},} \nb\\
{g_{{ {Y}}3}} &\equiv& { \frac{\beta_1}{2 c_1-c_{13}  c_-} \frac{2 \Delta s}{c_V} \eta^{1/5}  e_Y^i V^i, } \nb\\
{g_{{ {Y}}4}} &\equiv& { \frac{\beta_1}{2 c_1-c_{13}  c_-} \frac{2 \Delta s}{c_V} \eta^{1/5} N^i V^i.} 
\eqn


\end{document}